\documentclass
[aps,prc,twocolumn,showpacs,showkeys,amsmath,floatfix,superscriptaddress]
{revtex4-1}
\usepackage{graphicx}
\usepackage{mathptmx}                
\usepackage{dcolumn}                 
\usepackage{bm}                      
\usepackage{sistyle} 			
\usepackage{eurosym} 			
\DeclareMathOperator{\e}{e}
\newcommand{\be}{\begin{equation}}
\newcommand{\ee}{\end{equation}}
\newcommand{\bea}{\begin{eqnarray}}
\newcommand{\eea}{\end{eqnarray}}

\usepackage{color}
\usepackage{soul,latexsym, amssymb, amsmath, multirow, mathrsfs, booktabs}
\usepackage{longtable}
\usepackage{rotating}

\begin{document}
\title{
 Analytical mass formula and nuclear surface properties in the ETF approximation
}
\author{Fran\c{c}ois Aymard}
\affiliation{CNRS and ENSICAEN, UMR6534, LPC, 14050 Caen c\'edex, France}
\author{Francesca Gulminelli}
\affiliation{CNRS and ENSICAEN, UMR6534, LPC, 14050 Caen c\'edex, France}
\author{J\'er\^ome Margueron}
\affiliation{Institut de Physique Nucl\'eaire de Lyon, Universit\'e Claude Bernard Lyon 1, \\IN2P3-CNRS, F-69622 Villeurbanne Cedex, France}
\date{Dated: \today}
\begin{abstract}
The problem of the determination of the nuclear surface and surface symmetry energy is addressed in the framework of the Extended Thomas Fermi
(ETF) approximation using Skyrme functionals. 
We propose an analytical model for the density profiles with variationally determined diffuseness parameters.
For the case of symmetric nuclei, the resulting ETF functional can be exactly integrated, leading to an analytical formula 
 expressing the surface energy as a function of the couplings of the energy functional.  
The importance of non-local terms is stressed, which cannot be simply deduced from the local part of 
the functional.
In the case of asymmetric nuclei, we propose an approximate expression for the diffuseness and the surface energy.
These quantities are analytically related to the parameters of the energy functional. In particular, the influence of the different equation of state parameters can be explicitly quantified. Detailed analyses of the different energy components (local/non-local, isoscalar/isovector, surface/curvature and higher order) are also performed.
Our analytical solution of the ETF  integral
improves over previous models and leads to a 
precision better than $200$~keV per nucleon in the determination
of the nuclear binding energy for dripline nuclei.    
\end{abstract}
\maketitle
%
%
%
\section{Introduction}

Skyrme functionals have been widely used to describe nuclear structure properties, with different level of sophistication in the many-body treatment, 
from the simplest Thomas-Fermi~\cite{Brack1985} to modern multi-reference calculations~\cite{bender}.
The most basic observable accessible to the functional treatment is given by nuclear mass,  allowing the analysis of the different mass components in terms of bulk and surface properties, as well as isovector and isoscalar properties.
The theoretical prediction  of nuclear mass is not only important in itself, but it is also  a fundamental tool to optimize the different functional forms and associated parameters, 
for an increasing predictive power of density functional calculations \cite{goriely}. Indeed mass predictions from microcopic density functionals nowadays starts to equalize the most precise 
phenomenological mass formulas available in the literature\cite{ws3,moller_nix,duflo_zucker}.

For practical applications in nuclear structure or nuclear astrophysics problems, different parametrizations of nuclear masses fitted on density functional calculations with Skyrme forces have been proposed \cite{ETFSI, dan,mekjian,pei}. The limitation of these works is that the different coefficients are not analytically calculated but they result from the fit to the numerically determined nuclear masses. As a consequence, the fit has to be performed again each time that the functional is improved by adding further constraints from the rapidly improving experimental data. Moreover, the absence of an analytical link between the Skyrme parameters and the coefficients of the mass formula implies that it is difficult to make an unambiguous correlation between the different parts of the mass functional and the physical properties of the effective interaction.
For these reasons, it appears interesting to  search for an analytical expression of the mass formula coefficients, directly linked to the functional form and parameters of the Skyrme interaction. The derivation of such an analytical formula is the purpose of this paper.

An especially appealing formalism when seeking for analytical expressions is the semi-classical  Extended-Thomas-Fermi (ETF) approach, which is 
based on an expansion in powers of $\hbar$ of the energy functional~\cite{krivine,Centelles1990,etf_recent,chamel_etf,mekjian}.
The  advantage of the   ETF approximation is that the 
non-local terms in the energy density functional are entirely replaced by local gradients.
As a consequence, the energy functional solely depends on the local particle densities. 
Thus, the energy of any arbitrary nuclear configuration can be calculated if the neutron and proton density profiles $\rho_n$ and $\rho_p$ are given through a parametrized form.  
These density profiles are those of the ground state, or of any arbitrary excited state. A large number of configurations can therefore be explored, and this 
appealing property of ETF has been used to study nuclear configurations in dilute stellar matter contributing to the sub-saturation finite temperature equation of state
~\cite{avancini,pais,aymard}.
On the other side,
the well known limitation of ETF is that only the smooth part of the nuclear mass can be addressed, and shell effects have to be added on top, for instance through the well known Strutinsky integral theorem\cite{pearson} . 

In this paper,
we will consider an ETF expansion up to the second $\hbar^2$-order, and limit ourselves  to the smooth part of the mass functional.

The plan of the paper is as follows.
 
Section \ref{sec_sym_nucl_vacuum} addresses the problem of symmetric nuclei. A single density profile is supposed for protons and neutrons, and symmetry breaking effects are included by accounting for the Coulomb modification of the bulk compressibility. In this simplified case, 
the ETF integrals can be analytically integrated leading to a very transparent form for the surface and curvature terms of the nuclear energy 
(section \ref{sec_sym_energy}) and of the surface diffuseness (section \ref{sec_sym_diffuseness}) . In this same section we also retrive (section \ref{sec_sym_results}) that in a one-dimensional geometry 
the local and non-local terms are related, and the surface tension can be consequently be expressed as a function of the local terms only~\cite{krivine}. 
This means that the
surface tension solely depends on the local components of the energy density functional, that is the bulk properties of nuclear matter, and it does not depend on the non-local gradient and spin-orbit terms. 
This remarkable property however breaks down in spherical symmetry, and any, even slight approximation to the exact variational profile, for instance the use of parametrized densities,
increases  the difference between local and non-local contributions to the surface energy. As a consequence, using parametrized density profiles, the contribution of non-local terms has to be carefully calculated independently of the local part, and 
the two separate contributions must be summed up to obtain the surface energy and the surface tension.
 
The more general problem of isospin asymmetric nuclei is studied in section \ref{sec_asym_nucl_vacuum}.
We first demonstrate in section \ref{sec_asym_changing_sym} that a large part of the isospin dependence can be accounted for, if the  asymmetry dependence of the saturation density is introduced in the nuclear bulk. The residual surface symmetry part  is then defined in terms of the  isovector density. This energy density term  is not analytically integrable, meaning that approximations have to be performed.
We propose in section \ref{sec_asym_Gauss_apprx} two different approximations and critically discuss their validity in comparison both to numerical integration of the ETF functional, and to complete Hartree-Fock (HF) calculations using the same functional (section \ref{sec_IV_compHF}). The first approximation, inspired from Ref.~\cite{krivine_iso}, consists in neglecting the neutron skin (section \ref{sec_asym_TK}). Surprisingly enough, this very crude approximation leads to an overestimation of Hartree-Fock energies of medium-heavy and heavy nuclei of no more than 200-400 
keV/nucleon even for the most extreme dripline nuclei. Again, such an accuracy can be obtained only if both local and non-local terms in the energy functional are separately calculated, meaning that the symmetry energy does not only depend on bulk nuclear matter properties. This might be at the origin of the well known ambiguities in the extraction of the symmetry energy from density functional calculations of finite nuclear properties~\cite{pei,dan03,reinhardt,douchin}.
A better accuracy  for neutron rich nuclei is obtained if isospin fluctuations are accounted for, and in section \ref{sec_asym_gauss_tot} the approximation is made that the surface symmetry energy density is strongly peaked at the nuclear surface. 

Finally the complete mass formula is calculated for different representative Skyrme functionals in section \ref{seq_applications}.
The qualitative behavior of the different energy components,
that is the surface, curvature and higher order terms decomposed into 
isovector and isoscalar parts, and local and non-local parts,
is discussed.
The different analytical expressions for the mass functional are explicitly demonstrated in the appendix  and can be  readily used with any Skyrme interaction.
The paper is summarized in section \ref{sec_conclu}, and conclusions are given. 

\section{Symmetric nuclei}
\label{sec_sym_nucl_vacuum}
Let us first consider a locally symmetric matter distribution, that is characterized by a single density profile which is supposed to be identical for protons and neutrons.  
This idealized situation is not completely realistic even in $N=Z$ nuclei because of  isospin symmetry breaking due to Coulomb.
However, it was shown~\cite{aymard} that a great part of the Coulomb isospin symmetry breaking effects can be included simply 
accounting for the Coulomb modification of the bulk compressibility~\cite{ldm,centelles1,centelles2}. 
This  single-density model leads to an excellent reproduction of 
the microscopically calculated as well as experimentally measured magic $N=Z$ nuclei over the periodic table~\cite{aymard,pana}.

The idealized case of a common density profile
for protons and neutrons has  the advantage of leading to exact formulas for the nuclear binding energy, as we explicitly show in this section. 
As we will see, 
this allows disentangling in a non-ambiguous way bulk, surface, curvature as well as higher order terms, and to determine exact relations 
connecting the different energy components to the parameters of the 
energy functional.

Neglecting spin-gradient terms, the ETF Skyrme energy density  reads, 
\bea
\mathcal{H}[\rho] =
h(\rho) 
+ \frac{\hbar^2}{2m^*(\rho)}  \tau_2 
+ \left ( C_{fin} - \frac{C_{so}^2}{\hbar^2}  \rho m^*(\rho)  \right )
 \left( \boldsymbol{\nabla}\rho \right)^2.\nonumber \\
\label{eq_sym_density_energy_Skyrme}
\eea
In this expression, $m^*(\rho)$ is the density dependent effective mass, $m/m^*=1+\frac{2m}{\hbar^2}C_{eff}\rho$, 
the kinetic energy density consists of a  zero order Thomas-Fermi term $\tau_0$
as well as of a second order local and non-local correction $\tau_2=\tau_2^l + \tau_2^{nl}$:
\bea
\tau_0      & =& \frac{3}{5} \left( \frac{3\pi^2}{2} \right)^{2/3} \rho^{5/3}  \\
\tau_2^l    & =& \frac{1}{36} \frac{ \left( \boldsymbol{\nabla}\rho \right)^2 }{\rho} + \frac{1}{3} \Delta\rho  \label{tau_loc}\\
\tau_2^{nl} & =&
\frac{1}{6} \frac{ \boldsymbol{\nabla}\rho \boldsymbol{\nabla} f}{f}
+\frac{1}{6} \rho  \frac{\Delta f}{f}
- \frac{1}{12} \rho  \left(  \frac{\boldsymbol{\nabla} f}{f} \right) ^2, \label{tau_nloc}
\eea
with $f=m/m^*$. The local terms are given by:
\bea
h(\rho) = 
\frac{\hbar^2}{2m^*(\rho)}  \tau_0 + C_0\rho^2 + C_3\rho^{\alpha + 2},
\label{eq_energy_zeroth_loc_sym} 
\eea
and gradient terms arise both from the non-local and the spin-orbit part of the Skyrme functional.
$(C_0,C_3,C_{eff},C_{fin},C_{so},\alpha)$ are Skyrme parameters, given in appendix \ref{sec_effective_interaction}. Spin-gradient terms are not considered in the 
applications of this paper, but their inclusion is straightforward. Full expressions are given in appendix~\ref{sec_effective_interaction}.   We will also limit
ourselves to spherical symmetry throughout the paper.
 
To compute Eq.~(\ref{eq_sym_density_energy_Skyrme}), the density profile $\rho(r)$ is required.
The most common choice in the literature \cite{pana} consists in taking a Fermi function. In particular, 
it was shown~\cite{aymard} that a Fermi function succeeds in well reproducing 
the density profiles and the corresponding energy
calculated with the spherical HF model. The density profile reads, 
\bea
\rho(r) = \rho_{sat} F(r)
\;\; ; \;\;
F(r) = \left( 1+\e^{ (r-R)/a } \right)^{-1}
.
\label{eq_sym_density_profile}
\eea
In this equation, $\rho_{sat}$ is the saturation density of symmetric nuclear matter, and $R$ is the radius parameter related to the  particle number of the nucleus
\bea
A \simeq
\frac{4}{3} \pi \rho_{sat} R^3 	\left[
										1 +
										\pi^2 \left( \frac{a}{R} \right)^2
								\right]
\label{eq_sym_3D_A}
.
\eea
Let us observe that Eq. (\ref{eq_sym_3D_A}) is a finite expansion and does not require any assumption
except that $\e^{-R/a} \ll 1$, that is $a \lesssim R$ \cite{krivine_math}. 
If, in addition, we assume $a \ll R$, we can invert equation (\ref{eq_sym_3D_A}) 
to get at the fourth order in $(a/R)$
\bea
R =
R_{HS} 
\left[
		1 - \frac{\pi^2}{3} 	\left( \frac{a}{R_{HS}} \right)^2 
			+ O\left( \left(\frac{a}{R_{HS}}\right)^4  \right)
\right]
\label{eq_sym_3D_R_1}
,
\eea
where $R_{HS} = A^{1/3}r_{sat}$ is the equivalent homogeneous sphere radius, and
$r_{sat}= \left(\frac{4}{3}\pi\rho_{sat}\right)^{-1/3}$ is the mean radius per nucleon.

The two other parameters entering Eq.~(\ref{eq_sym_density_profile}) are
the diffuseness $a$ of the density profile,  which is analytically derived in section~\ref{sec_sym_diffuseness},
and the saturation density $\rho_{sat}$ which corresponds 
to the equilibrium density  of homogeneous infinite symmetric matter.

\subsection{Ground state energy}
\label{sec_sym_energy}
Integrating in space Eq.~(\ref{eq_sym_density_energy_Skyrme})
computed with the parametrized density profile given by Eq.~(\ref{eq_sym_density_profile}),
allows obtaining the total energy $E$ of a nucleus of a mass $A$ defined by Eq.~(\ref{eq_sym_3D_A}):
\bea
E =
\int_0^{\infty} \mathrm d\mathrm{\mathbf{r}} \mathcal{H}[\rho(\mathrm{r})] .
\label{eq_sym_Etot}
\eea
Within the nucleus total binding energy, it is interesting to distinguish
the bulk, surface, and curvature components corresponding to different functional dependences on the nuclear 
size~\cite{dudek}. 

The bulk energy $E_{b}$ is the energy of a finite volume of nuclear matter. It corresponds 
to the energy that  the nucleus would have without
finite-size effects, defined by:
\bea
E_b = \mathcal{H}_{sat} V_{HS} = \lambda_{sat} A
,
\label{eq_sym_bulk}
\eea
where $V_{HS}=4/3\pi R_{HS}^3=A/\rho_{sat}$ is the equivalent homogeneous sphere volume and
$\lambda_{sat}$ is the energy per particle at saturation:  
\bea
\lambda_{sat}
=
\left. \frac{\mathcal{H}}{\rho} \right|_{\rho_{sat}}
= C_{kin}\frac{m}{m^*_{sat}} \rho_{sat}^{2/3}
+ C_0\rho_{sat}
+ C_3\rho_{sat}^{\alpha + 1}
,
\label{eq_lambda_sym_nm}
\eea
with
$C_{kin}=  \frac{3}{5}\hbar^2/(2m)( 3\pi^2/2 )^{2/3} $ and $m^*_{sat}=m^*(\rho_{sat})$.
The finite-size correction to the bulk energy   $E_{s}$
is defined as the total energy after the bulk is removed, that is
\begin{align}
E_{s} & =
\int \mathrm d\mathrm{\mathbf{r}} \mathcal{H}[\rho(\mathrm{r})]
- \mathcal{H}_{sat} V_{HS}
\nonumber \\ & =
4\pi \int_0^{\infty} \mathrm dr \big\{  \mathcal{H}[\rho(r)] - \lambda_{sat} \rho(r)  \big\} r^2
.
\label{eq_sym_3D_def_Enb}
\end{align}
This finite size contribution $E_{s}$  will be called the surface energy in the following.
Let us notice that if we have properly removed the bulk energy part by Eq.~(\ref{eq_sym_bulk}),
the  surface energy should scale with $A$ with a dependence slower than linear, but the dependence can be different from 
$A^{2/3}$ because of curvature and higher order terms, see also ref~.\cite{dudek}.
We will see in the following that it is indeed the case in spherical symmetry.
\\
In the energy density $\mathcal{H}[\rho]$ Eq.~(\ref{eq_sym_density_energy_Skyrme}),
we can distinguish the non-local terms which depend on the density derivatives and are pure finite-size effects,
from the local energy part $h(\rho)$ which only depends on the equation of state and on the density profile.
We then write the  surface energy as $E_s=E_ s^L+E_ s^{NL}$, 
with $E_ s^L$ the local part and $E_ s^{NL}$ the non-local one 
\cite{footnote1}:
\bea
E_ s^L & =& 4\pi
\int_{0}^{\infty} \mathrm dr \left\{  h[\rho(r)] - \frac{ h(\rho_{sat}) }{ \rho_{sat} } \rho(r)  \right\}  r^2,
\label{eq_sym_3D_Enb_L}
\\
E_ s^{NL} & = & 4\pi
\int_{0}^{\infty} \mathrm dr	
\left\{  
	\frac{\hbar^2\tau_2}{2m^*}
	+ \left ( C_{fin} 	- \frac{C_{so}^2}{\hbar^2} \rho m^* \right )
	(\nabla\rho)^2									
\right\} r^2 . \nonumber \\
\label{eq_sym_3D_Enb_NL}
\eea
To obtain Eq.~(\ref{eq_sym_3D_Enb_NL}), we have changed the Laplace derivatives into gradients 
in the kinetic part, see Eqs.~(\ref{eq_sym_rho_tau}),  integrating by parts.   
Making a simple variable change,  the originally $3$-dimensional integral can be turned into the sum of 
three $1$-dimensional integrals (see appendix~\ref{sec_appendix_int_FD}).
Then a very accurate approximation, that is with an error less than $\big(\exp(-5a/3R)-\exp(-a/R)\big)$, allows to analytically integrate the differences of Fermi functions,
such that the local and non-local terms can be written as a function of the effective interaction parameters as 
(calculation details are given in appendix~\ref{sec_appendix_sym_non_bulk}):
\bea
E_{s}^{L} & = & \mathcal{C}^{L}_{surf}   \frac{a(A)}{r_{sat}} A^{2/3} \nonumber \\ 
 &+& \mathcal{C}^{L}_{curv}   \left( \frac{a(A)}{r_{sat}} \right)^2 A^{1/3}\nonumber \\
 &+& \mathcal{C}^{L}_{ind}   \left( \frac{a(A)}{r_{sat}} \right)^3 \nonumber \\
 &+&  o \left(  \left( \frac{a(A)}{r_{sat}} \right)^4 A^{-1/3} \right);   \label{eq_sym_3D_Es_cl_L}
\eea
\bea
E_{s}^{NL} & = & \frac{1}{a^2(A)} \mathcal{C}^{NL}_{surf}    \frac{a(A)}{r_{sat}} A^{2/3} \nonumber \\
 &+& \frac{1}{a^2(A)} \mathcal{C}^{NL}_{curv}   \left( \frac{a(A)}{r_{sat}} \right)^2 A^{1/3} \nonumber \\
 &+&  \frac{1}{a^2(A)} \mathcal{C}^{NL}_{ind}    \left( \frac{a(A)}{r_{sat}} \right)^3 \nonumber \\
 &+&  o \left(  \left( \frac{a(A)}{r_{sat}} \right)^4 A^{-1/3} \right), \label{eq_sym_3D_Es_cl_NL}
\eea
where the coefficients $\mathcal{C}^{L(NL)}_{surf(curv)(ind)}$ depend on the 
saturation density $\rho_{sat}$ and on the Skyrme parameters
$ C_0, C_3, C_{eff}, \alpha, C_{fin}, C_{so}$,
and where we have anticipated the  (slight) $A$-dependence of the diffuseness in the most general case (see section~\ref{sec_sym_diffuseness}).

The coefficients $\mathcal{C}_{i}^{L}$ and  $\mathcal{C}_{i}^{NL}$ corresponding to the local and non-local energy components 
read (see appendix~\ref{sec_appendix_sym_L}):
\begin{widetext}
\bea
\mathcal{C}^{L}_{surf}  & = &
3
\left\{
	 C_{kin}\rho_{sat}^{2/3}
				\left[
					 \eta^{(0)}_{5/3} \frac{m}{m^*_{sat}}  
					 - \frac{3}{5}   \delta m_{sat} 
				\right]
	- C_0 \rho_{sat}
	+ C_3 \rho^{\alpha+1}_{sat}  \eta^{(0)}_{\alpha+2}
\right\} ,
\label{eq_sym_coeff_L_surf}
 \\
\mathcal{C}^{L}_{curv}  &  =&
6
\left\{
	 C_{kin}\rho_{sat}^{2/3}
				\left[
				 	  \left( \eta^{(1)}_{5/3} - \frac{\pi^2}{6} \right)  \frac{m}{m^*_{sat}} 
				 	- \frac{3}{5} \eta^{(0)}_{5/3}   \delta m_{sat}	
				\right]	
+ C_3 \rho_{sat}^{\alpha+1} 
	\left(
		\eta^{(1)}_{\alpha+2}
		- \frac{\pi^2}{6}
	\right)
\right\} ,
\label{eq_sym_coeff_L_curv}
 \\
\mathcal{C}^{L}_{ind}  & =&
3
\left\{
	 C_{kin}\rho_{sat}^{2/3}
		    	\left[	
					\left( \eta^{(2)}_{5/3} - \frac{2 \pi^2}{3} \eta^{(0)}_{5/3}  \right) \frac{m}{m^*_{sat}} 
					- \frac{2}{5}
						\left( 3\eta^{(1)}_{5/3} - \pi^2\right)  \delta m_{sat}
			\right]
	+ \frac{\pi^2}{3} C_0 \rho_{sat}
	+ C_3 \rho_{sat}^{\alpha+1} 
		\left(	
			\eta^{(2)}_{\alpha+2}
			- \frac{2 \pi^2}{3} \eta^{(0)}_{\alpha+2}
		\right)
\right\} ,
\label{eq_sym_coeff_L} \\
\mathcal{C}^{NL}_{surf} & =&
3
\left\{
	\frac{\hbar^2}{2m} \frac{1}{6} 
		\left(	
			\frac{1}{12}	
			- \frac{11}{36} \delta m_{sat}
			- \frac{1}{2} \sum_{i=0}^{i_{max}} (-1)^i  \frac{ (  \delta m_{sat} )^{i+2} }{ (i+3)(i+4) }
		\right)
	+ \frac{1}{6} C_{fin} \rho_{sat}
	+ V_{so} \rho_{sat}^2  \sum_{i=0}^{i_{max}} (-1)^i  \frac{ (  \delta m_{sat} )^i }{ (i+3)(i+4) }
\right\} ,
\label{eq_sym_coeff_NL_surf}
 \\
\mathcal{C}^{NL}_{curv} & =&
6
\left\{
	\frac{\hbar^2}{2m} \frac{1}{6} 
		\left(	
			\frac{1}{12}	
			- \frac{1}{2}   \sum_{i=0}^{i_{max}} (-1)^i  \frac{ ( \delta m_{sat} )^{i+2} }{ (i+3)(i+4) }
					\left[  \eta^{(0)}_{i+2} + 1  \right]
		\right)
	+ V_{so} \rho_{sat}^2  \sum_{i=0}^{i_{max}} (-1)^i  \frac{ ( \delta m_{sat} )^i }{ (i+3)(i+4) }
			\left[  \eta^{(0)}_{i+3} + 1  \right]
\right\} ,
\label{eq_sym_coeff_NL_curv}
 \\
\mathcal{C}^{NL}_{ind}  & =&
6
\left\{ 
	\frac{\hbar^2}{2m} \frac{1}{6} 
		\left(	
			- \frac{1}{12} \frac{\pi^2}{6}
			+ \frac{11}{36}\left( 1+\frac{\pi^2}{6} \right) \delta m_{sat}
			- \frac{1}{2} \sum_{i=0}^{i_{max}} (-1)^i  \frac{ ( \delta m_{sat} )^{i+2} }{ (i+3)(i+4) }
					\left[  \eta^{(1)}_{i+2} + \eta^{(0)}_{i+2} - \frac{\pi^2}{3}  \right]
		\right) \right. \nonumber \\
	&-& \left. \left( 1+\frac{\pi^2}{6} \right) C_{fin} \rho_{sat}
	+6 V_{so} \rho_{sat}^2  \sum_{i=0}^{i_{max}} (-1)^i  \frac{ ( \delta m_{sat} )^i }{ (i+3)(i+4) }
			\left[  \eta^{(1)}_{i+3} + \eta^{(0)}_{i+3} - \frac{\pi^2}{3}  \right] \right\} .
\label{eq_sym_coeff_NL}
\eea
\end{widetext}
 where $\delta m_{sat}=(m-m^*_{sat})/m^*_{sat}$, $V_{so}=-m C_{so}^2/\hbar^2$, 
and where we have introduced the coefficients $\eta_\gamma^{(k)}$
defined by equation~(\ref{eq_appendix_gen_eta}).
Their numerical values are given in the same appendix.
%
In order to have an analytical expression, we have made in eqs.~(\ref{eq_sym_coeff_NL_surf})-(\ref{eq_sym_coeff_NL})
a Taylor expansion of the effective mass inverse
$f^{-1} = \sum_{i=0}^\infty (-1)^i ( \delta m  )^i$. 
This expansion is rapidly convergent: a truncation at $i_{max}=7$ produces an error
 $\sim 1\%$ at the highest  possible density $\rho_{sat}=0.16$ fm$^{-3}$ 
in the case of the SLy4 interaction.


Equations~(\ref{eq_sym_3D_Es_cl_L}) and~(\ref{eq_sym_3D_Es_cl_NL}) show that the dominant surface effect in the symmetric nucleus energetics  is, as expected, a term $\propto A^{2/3}$.
As it is well known, this term fully exhausts the finite-size effects given by the presence of a nuclear surface in the one-dimensional case of a semi-infinite slab geometry~\cite{krivine,dan}.
Indeed in this case we have, see appendix~\ref{sec_appendix_3Dto1D},
\bea
E_{s}^{slab} =
\int_{-\infty}^{+\infty} \mathrm dx \big\{  \mathcal{H}[\rho(x)] - \lambda_{sat} \rho(x)  \big\} .
\label{eq_sym_def_Enb_SLAB} 
\eea
The evaluation of the integral Eq.~(\ref{eq_sym_def_Enb_SLAB})
 leads to the same $\propto A^{2/3}$  term as in the spherical geometry, with a modified form factor $4\pi R_{HS}^2$:
\bea
 \sigma =
\sigma^{L} + \sigma^{NL}
=
\left(
	\mathcal{C}^{L}_{surf}   +
	\frac{1}{a^2} \mathcal{C}^{NL}_{surf} 
\right)
\frac{a}{4\pi r_{sat}^3}
.
\label{eq_sym_Enb_SLAB}
\eea
where $\sigma = \lim_{A\to \infty}E_{s}^{slab}/A^{2/3} $ is the slab surface tension. 
The form factor difference between the surface energy of the slab and the one in spherical symmetry
signs the difference of geometry, and
the spherical surface energy is the surface area multiplied 
by the energy per unit area of the infinite tangent plane.
Let us notice that since the mass cannot be defined in the semi-infinite medium, 
the diffuseness in Eq.~(\ref{eq_sym_Enb_SLAB}) is a constant.

In a three-dimensional geometry, the existence of a surface   leads to additional finite-size terms, even in the spherically symmetric case, as shown by  eqs.~(\ref{eq_sym_3D_Es_cl_L}),~(\ref{eq_sym_3D_Es_cl_NL}).
The terms proportional to $A^{1/3}$ are the so-called curvature terms which
correct the surface energy with respect to the slab tangent limit.
It is interesting to notice that we also have $A$-independent terms, which are rarely accounted for in the literature but turn out to be important for light nuclei~\cite{dudek}.
As it can be seen in Eq.~(\ref{eq_sym_3D_R_1}),  higher order terms are of the order $\propto A^{-1/3}$ and are systematically neglected in this work.
This Taylor expansion is known in the literature as the leptodermous expansion~\cite{Myers1985,dudek}.
It is interesting to observe that both local and non-local plane surface, curvature, and mass independent energy components arise even if no explicit gradient term is included in the functional.
As a consequence, surface properties are determined by a complex interplay between equation of state properties and specific finite nuclei properties like spin-orbit
and finite range. 
Using the definitions of 
the energy per particle at saturation 
$\lambda_{sat}= \left. h/\rho \right|_{\rho_{sat}} = \left. \partial h/\partial\rho \right|_{\rho_{sat}}$
and the nuclear symmetric matter incompressibility $K_{sat}=9\rho_{sat}^2\partial^2 (h/\rho)/\partial\rho^2|_{\rho_{sat}}$,
we can express the local energy eqs.~(\ref{eq_sym_coeff_L_surf}),~(\ref{eq_sym_coeff_L_curv}) and (\ref{eq_sym_coeff_L})
as a function of nuclear matter properties only, using the following expressions:
\begin{widetext}
\bea
\alpha & = &
- \frac{
		K_{sat} + 9 \lambda_{sat} - C_{kin}\rho_{sat}^{2/3} \left( \frac{m}{m^*_{sat}} + 3 \delta m_{sat} \right)
	}
	{
		9 \left[ \lambda_{sat} - C_{kin}\rho_{sat}^{2/3} \left( \frac{m}{3 m^*_{sat}} - \delta m_{sat} \right)   \right]
	} ,
\label{eq_sym_alpha} \\
C_0\rho_{sat} & = &
\frac{
		\lambda_{sat}K_{sat} - C_{kin}\rho_{sat}^{2/3} K_{sat} \frac{m}{m^*_{sat}} 
		- C_{kin}\rho_{sat}^{2/3} \lambda_{sat} \left( 4 \frac{m}{m^*_{sat}} + 21 \delta m_{sat} \right) - 9 C_{kin}^2\rho_{sat}^{4/3} \delta m_{sat}
	}
	{
		K_{sat} + 9 \lambda_{sat} - C_{kin}\rho_{sat}^{2/3} \left( \frac{m}{m^*_{sat}} + 3 \delta m_{sat} \right)
	} ,
\label{eq_sym_C0} \\
C_3\rho_{sat}^{\alpha+1} & = &
\frac{
		9 \left[ \lambda_{sat} - C_{kin}\rho_{sat}^{2/3} \left( \frac{m}{3 m^*_{sat}} - \delta m_{sat} \right)   \right]^2	
	}
	{
		K_{sat} + 9 \lambda_{sat} - C_{kin}\rho_{sat}^{2/3} \left( \frac{m}{m^*_{sat}} + 3 \delta m_{sat} \right)
	}
	.
\label{eq_sym_C3}
\eea
\end{widetext}

The expression of the coefficients $\mathcal{C}_{i}^{(N)L}$ greatly simplifies if we consider a simplistic Zamick-type interaction,
with $\alpha = 1$ and $m=m^*$:
\bea
\mathcal{C}^{L}_{surf} & =&
\left( \frac{9}{5} \eta_{5/3}^{(0)} + \frac{3}{2} \right) e_{sat}^F - \frac{3}{2} \lambda_{sat} ,
\label{eq_sym_L_surf_Zam} \\
\mathcal{C}^{NL}_{surf} & =&
\frac{1}{24} \frac{\hbar}{2m} + \frac{1}{2} C_{fin} \rho_{sat} ,
\label{eq_sym_NL_surf_Zam},
\eea
where we have introduced the Fermi energy per nucleon at saturation $e^F_{sat}=\frac 5 3 C_{kin}\rho_{sat}^{2/3}$.
 
We can see that even in this oversimplified model the nuclear surface properties cannot be simply reduced to EoS parameters.
We can also gather the local and non-local terms in order to classify finite-size effects
according to the rank of the Taylor expansion.
Thus we introduce the surface $E_{surf}$, curvature $E_{curv}$ and $A$-independent $E_{ind}$ energy components:
\begin{align}
E_{surf} &
= \left[ \mathcal{C}^{L}_{surf} 
+ \frac{1}{a^2(A)} \mathcal{C}^{NL}_{surf}  \right] \frac{a(A)}{r_{sat}} A^{2/3} ,
\label{eq_sym_3D_Es_cl_surf}
\\
E_{curv} &
= \left[ \mathcal{C}^{L}_{curv} 
+ \frac{1}{a^2(A)} \mathcal{C}^{NL}_{curv}\  \right] \left( \frac{a(A)}{r_{sat}} \right)^2 A^{1/3} ,
\label{eq_sym_3D_Es_cl_curv}
\\
E_{ind} & 
= \left[ \mathcal{C}^{L}_{ind} 
+  \frac{1}{a^2(A)} \mathcal{C}^{NL}_{ind}  \right]  \left( \frac{a(A)}{r_{sat}} \right)^3
\label{eq_sym_3D_Es_cl_ind}
.
\end{align}
We can see that all terms are multiplied by a power of the diffuseness except the non-local curvature part which is not.
The role of the diffuseness on the surface properties thus depends on the rank of the Taylor expansion (surface, curvature, independent,...)
and is not the same for  the local or the non-local part.
The functional difference between the local and non-local terms
comes from the squared density gradient appearing in the non-local energy Eq.~(\ref{eq_sym_3D_Enb_NL}).
Globally, if the  diffuseness is high, the local energy dominates over the non-local one, see eqs.~(\ref{eq_sym_3D_Es_cl_surf})-(\ref{eq_sym_3D_Es_cl_ind}). 
This is easy to understand: in the limit 
of a purely local energy functional, the optimal configuration corresponds to a homogeneous hard sphere at saturation density, given by $a=0$.
The existence of a  finite diffuseness for atomic nuclei is due to the presence of non-local terms in the functional, 
because of both explicit gradient interactions and of quantum effects on the kinetic energy density.
Let us notice that both effects are present even in the simplified eqs.~(\ref{eq_sym_L_surf_Zam}),~(\ref{eq_sym_NL_surf_Zam}).

\subsection{Analytical expression for the diffuseness}
\label{sec_sym_diffuseness}
The ground state energy of this model for symmetric nuclei is given by the minimization of the energy per nucleon $\delta(E/A)=0$
with the constraint of a given mass number $A$.
We have seen in section~\ref{sec_sym_energy} that  
 the only unconstrained parameter of the model is the diffuseness parameter $a$.
 Though it does not play any role in the bulk energy, 
it is an essential ingredient for the surface energy $E_{s}$
given by eqs.~(\ref{eq_sym_3D_Es_cl_L}), (\ref{eq_sym_3D_Es_cl_NL}).
The diffuseness parameter can therefore be obtained from the variational equation~\cite{krivine}: 
\bea
\frac{\partial E_{s}}{\partial a} =0
\label{eq_sym_diffuseness_min_E_a}
.
\eea
In principle, one should also add the surface Coulomb energy into $E_s$,
which would change the variational equation. 
However, the resulting correction on $a$ is very small~\cite{pana}.

Equation~(\ref{eq_sym_diffuseness_min_E_a}) turns out to be particularly simple in the one-dimensional case of semi-infinite 
 matter, or equivalently neglecting curvature and $A$-independent terms when considering nuclei.
 Indeed, in this case, Eq.~(\ref{eq_sym_Enb_SLAB}) leads to an analytical solution, already obtained in Ref.~\cite{krivine}:
\begin{align}
a & = 
\sqrt{ 
		\frac{ \mathcal{C}^{NL}_{surf}}{ \mathcal{C}^{L}_{surf} } 
	}
.
\label{eq_sym_diffuseness_SLAB}
\end{align}
This equation shows that 
the slab diffuseness $a$ is determined by the balance between the local terms, which favour low diffuseness values
corresponding to a hard sphere of matter at saturation density; 
and non-local terms which favour a large diffuseness corresponding to matter close to uniformity.

The complete spherical case leads to the following $4^\mathrm{th}$ degree polynomial equation:
\bea
&3& \mathcal C_{ind}^{L}           \left( \frac{a}{r_{sat}} \right)^4
+
2 \mathcal C_{curv}^{L} A^{1/3}  \left( \frac{a}{r_{sat}} \right)^3 \label{eq_sym_diffuseness_3D} \\
&+&
\left( \mathcal C_{surf}^{L} A^{2/3} + \frac{1}{r_{sat}^2} \mathcal C_{ind}^{NL} \right)  \left( \frac{a}{r_{sat}} \right)^2
-
\frac{1}{r_{sat}^2} \mathcal C_{surf}^{NL} A^{2/3}
= 0
.
\nonumber
\eea
which has to be solved numerically.
\begin{figure}[htbp]
\includegraphics[width=0.9\linewidth, clip]{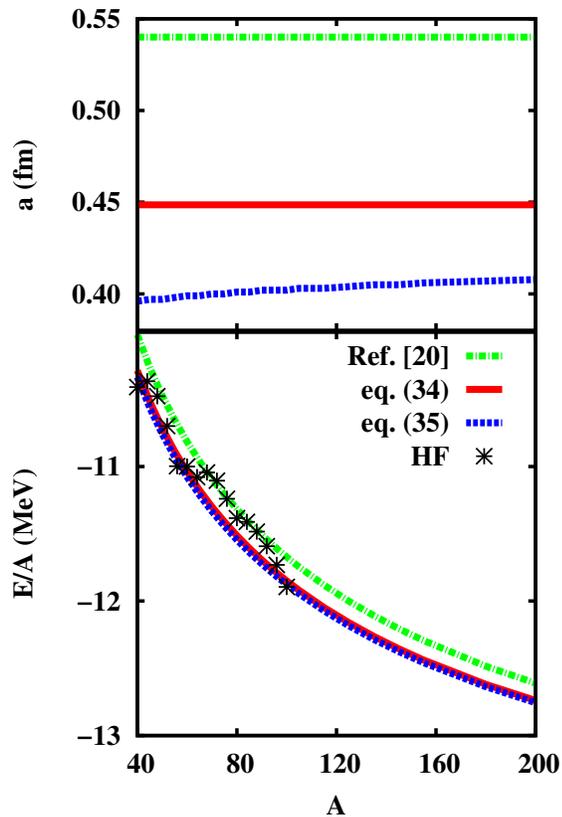}
\caption{
(Color online) Diffuseness (upper panel) and energy per nucleon (lower panel) of symmetric nuclei as a function of the mass number. 
Full red lines: calculations using the slab diffuseness Eq.~(\ref{eq_sym_diffuseness_SLAB}).
Dashed blue lines: calculations using the spherical diffuseness Eq.~(\ref{eq_sym_diffuseness_3D}).
Dash-dotted green lines: calculations using the diffuseness fitted from HF density profiles~\cite{pana}.
Star symbols: full Hartree-Fock calculations in spherical symmetry. 
}
\label{Fig_sym_diffuseness_isoscal}
\end{figure}

Notice that the coefficient $\mathcal C_{curv}^{NL}$ does not contribute to this equation since, as already mentioned, it does not depend on $a$, cf. Eq.(\ref{eq_sym_3D_Es_cl_curv}).
The solution of this equation, as well as the slab  approximation Eq.~(\ref{eq_sym_diffuseness_SLAB}), 
are shown in the case of the SLy4 interaction in the upper panel of Figure~\ref{Fig_sym_diffuseness_isoscal}.
We can see that the mass dependence of the diffuseness parameter $a$ in the general case is very small. This  agrees with the findings
 of Ref.~\cite{pana} (green dash-dotted lines), where 
the diffuseness parameter was extracted from a fit of Hartree-Fock density profiles.
Considering only the surface term we get $a\approx 0.45$ fm, while
we can observe that taking into account terms beyond surface (curvature and mass independent), 
the diffuseness is  shifted to lower values of the order of $a\approx 0.4$ fm.
This relatively large effect is due to the fact that the non-local curvature term does not contribute to the diffuseness (see Eq.~(\ref{eq_sym_3D_Es_cl_curv})).
Therefore  the effect of the curvature energy is to increase the local component, which tends to favor a low diffuseness.  

The energy per nucleon is shown in the lower panel of Fig.~\ref{Fig_sym_diffuseness_isoscal}, for the three models considered in the upper panel, and in comparison to HF calculations.
We can see from this figure that 
the variational approach systematically produces more binding than the use of a fitted value for the diffuseness, as 
we could have anticipated. Indeed the value of Ref.~\cite{pana} was obtained from a fit of the density, which does not guarantee a minimal energy. 
Less expected is the fact that the  energies calculated with the three different choices for the diffuseness  
are very close, though the value of the diffuseness  are quite different.
Specifically, implementing  the different diffusenesses into Eq.~(\ref{eq_sym_Etot}), the resulting total energy
reproduces the Hartree-Fock nuclear energies with very similar accuracy.

We can then conclude that introducing higher order terms in the variational derivation of the diffuseness, as it has been done in equation~(\ref{eq_sym_diffuseness_3D}), does not significantly improve the predictive power of the model.
 Therefore we will preferentially use the simpler expression of the slab diffuseness given by equation~(\ref{eq_sym_diffuseness_SLAB}). This choice is made in all the following figures, unless explicitly specified.

%
%
%
%
\subsection{Decomposition of the surface energy}
\label{sec_sym_results}
%
\begin{figure}[htbp]
\includegraphics[angle=270,width=\linewidth, clip]{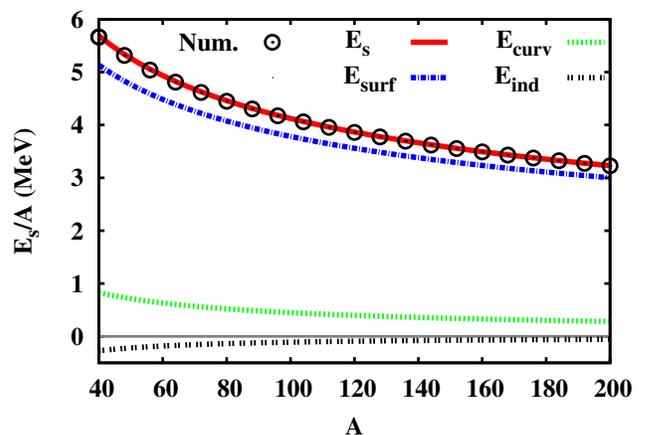}
\caption{
(Color online)
Numerical (black circles) and analytical (full red line) surface energy per nucleon (see text),
and its analytical decomposition into 
plane surface ( $\propto A^{2/3}$, dashed-dotted blue line), curvature ( $\propto A^{1/3}$, dotted green line),  and mass independent (double dotted black line) components
(eqs~(\ref{eq_sym_3D_Es_cl_surf}), (\ref{eq_sym_3D_Es_cl_curv}), (\ref{eq_sym_3D_Es_cl_ind}))
of symmetric nuclei, as a function of the mass number.
}
\label{Fig_sym_validity_analytical}
\end{figure}
 %
%
%
In this section, we study the functional behavior of the analytical formulas of section \ref{sec_sym_diffuseness}.
For these applications, we keep on focussing on a specific Skyrme interaction, namely SLy4~\cite{sly4}. 

In order to verify the accuracy of the analytical expression for the surface energy $E_{s}$,
we compare in Figure~\ref{Fig_sym_validity_analytical} the sum of eqs.
~(\ref{eq_sym_3D_Es_cl_surf})-(\ref{eq_sym_3D_Es_cl_ind})
with the numerical integration of Eq.~(\ref{eq_sym_3D_def_Enb}), as a function of the nucleus mass.
We can see  that the analytical expressions (full red line) very well reproduce the numerical values of $E_{s}$ (black circles).
An error smaller than $50$~keV per nucleon is obtained for the lightest considered nuclei, 
which rapidly vanishes with increasing  mass.
The deviation for light nuclei comes from  the approximation in the relation between the radius $R$ and the mass $A$.
Indeed, the expansion of the radius parameter Eq.~(\ref{eq_sym_3D_R_1})
leads to an expansion up to  $ A^{1/3} (a/r_{sat}) $ for $E_s$.
The missing terms $\propto  A^{-1/3} (a/r_{sat})^4 $ rapidly vanish with $A$, 
explaining the excellent reproduction of the exact numerical integral.

Figure~\ref{Fig_sym_validity_analytical} also
shows the plane surface, curvature and $A$-independent energy per nucleon components defined in
eqs.~(\ref{eq_sym_3D_Es_cl_surf}), (\ref{eq_sym_3D_Es_cl_curv}) and (\ref{eq_sym_3D_Es_cl_ind}).
Comparing the total surface energy $E_{s}$ (full red line) with $E_{surf}$ (dashed-dotted blue line), 
we can see that the $A^{2/3}$ dependence dominates over the whole mass table.
However, the curvature part (dotted green line), which represents the energetic cost of a spherical geometry, 
cannot be neglected even for heavy nuclei, impacting the total energy of $\gtrsim 300$~keV per nucleon for the heaviest nuclei.
For lighter nuclei ($A\lesssim 100$), the curvature contribution  to the total finite-size effects is of the order of $\sim 20\%$.
Though the $A$-independent energy (black dotted line) can be neglected from $A\gtrsim100$ for which $E_{ind}/A\lesssim50$~keV,
it should be taken into account for light nuclei if high accuracy is requested. 
Indeed, for $A=40$, the $A$-independent term contributes $\sim 5\%$ of the total surface energy.

\begin{figure}[htbp]
\includegraphics[angle=270,width=\linewidth, clip]{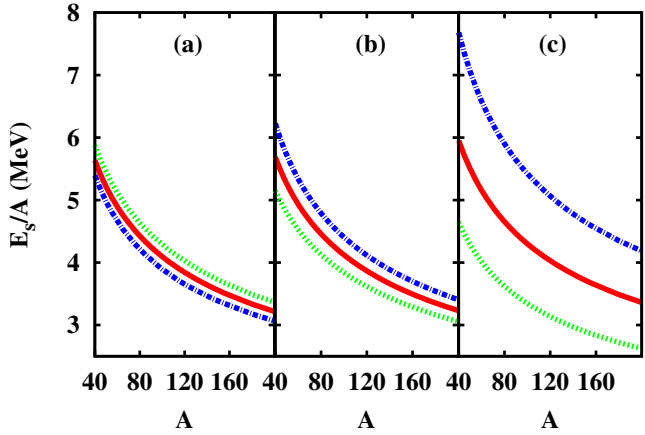}
\caption{
(Color online)
Surface energy per nucleon of symmetric nuclei using different choices for the diffusivity parameter $a$. Panel a): variational diffuseness including finite size effects from 
Eq.(\ref{eq_sym_diffuseness_3D}); panel b): variational diffuseness neglecting curvature terms from Eq.(\ref{eq_sym_diffuseness_SLAB}); panel c): diffuseness fitted from HF calculations in Ref.\cite{pana} $a=0.54$ fm. Red lines: total surface energy per nucleon.
Blue (green) lines: local (non-local) part multiplied by two.
}
\label{Fig_sym_Enb_decompo}
\end{figure}

 We now turn to the decomposition of the surface energy into a local and a non-local component.
It was shown in Ref.~\cite{krivine} that the local and non-local terms are expected to be exactly equal in the 
case of symmetric matter in a semi-infinite slab geometry. This result comes from the fact that the one-dimensional 
Euler-Lagrange variational equation can be solved by quadrature~\cite{wilets}. 
As a consequence,
it is easy to show that if the density profile is the exact solution of the Euler-Lagrange variational equation,
the first moment of the Euler-Lagrange equation implies that
the contribution of the local term in the surface energy density is at each point of space equal to the contribution of the non-local term, leading to the global equality between the local and non-local slab surface tensions:
\bea
\sigma^{L} = \sigma^{NL}. \label{magic} 
\eea
Extended to finite nuclei, this result would imply that
 only the local properties of the interaction (that is: the equation of state) are needed
to predict the surface properties of finite nuclei.

In this paper, we do not solve the Euler-Lagrange equation since we impose a given density profile,
but we do use a variational approach in minimising the energy to obtain the diffuseness parameter.
Therefore, it is easy to show that our model verifies the previous theorem in the one-dimensional case.
Indeed, using the slab diffuseness Eq.~(\ref{eq_sym_diffuseness_SLAB}), equation~(\ref{eq_sym_Enb_SLAB}) reads, 
\bea
\sigma^{L} = \sigma^{NL} = \lim_{A\rightarrow\infty}\frac{1}{2} \frac{E_{s}^{slab}}{A^{2/3}} = 
\frac{1}{4\pi r^3_{sat}}  \sqrt{ \mathcal{C}^{L}_{surf} \mathcal{C}^{NL}_{surf} }
.
\label{eq_sym_surf_diffuseness}
\eea
At first sight this result might look surprising since we have reduced the full variational problem to the variation of a single variable,
which represents a very poor variational approach. 
Equality (\ref{eq_sym_surf_diffuseness}) simply means that verifying the Euler-Lagrange first moment
is equivalent to minimising the energy with respect to a single free parameter.
That is, the density derivative is well described by the same parameter, here the diffuseness $a$,
as the density itself.

Unfortunately, this elegant theorem cannot be extended to the case of a spherical geometry.
Indeed, it is easy to show that the integrated Euler-Lagrange first moment leads to~\cite{thesis_aymard}
\bea
E_s^L - E_s^{NL}
=
4\int_0^\infty \mathrm dr
\int_\infty^r \mathrm dr' \frac{ \varepsilon_{NL}(r') }{r'}
.
\label{eq_theo_magik_3D}
\eea
The addition of this  non-zero integral to the local energy is due to the gradient 
part ($\propto 1/r$) of the spherical Laplacien, which comes from the difference between the plane and the spherical geometry, that is the 
spatial curvature. Eq.~(\ref{eq_theo_magik_3D}) shows that in a three-dimensional geometry
the equality between the local and non-local terms is violated for all components of the surface energy, including the term $\propto A^{2/3}$.

The left panel of Figure~\ref{Fig_sym_Enb_decompo} 
displays the decomposition of the surface energy between local (dashed-dotted blue line) and non-local (dotted green line) components, when the diffuseness of the density profile is consistently obtained from the numerical solution of the variational equation Eq.~(\ref{eq_sym_diffuseness_3D}).
We can see that the two terms are indeed different. 
This difference is however small, and the non-local energy only slightly dominates over the local one. This difference is amplified if the ansatz for the density profile deviates from the variational one. As an example, the central panel in Figure \ref{Fig_sym_Enb_decompo}
shows the surface energy obtained if the simpler expression Eq.~(\ref{eq_sym_diffuseness_SLAB}) for the diffuseness is employed.  The diffuseness extracted from a numerical fit of Hartree-Fock density profiles is employed following \cite{pana} in the right panel. We can see that the difference between local and non-local terms is increased as 
we consider density profiles increasingly deviating from the exact variational result. 

As we have already remarked, a higher diffusivity (from a) to c)) trivially leads to a globally higher surface energy. 
More interesting, the increased deviation from the exact variational result from a) to c)
 leads to a considerable increase of the local energy over the non-local one. 
This is a direct consequence of Eqs.~(\ref{eq_sym_3D_Es_cl_surf})-(\ref{eq_sym_3D_Es_cl_ind}).
 
\begin{figure}[htbp]
\includegraphics[angle=270,width=\linewidth, clip]{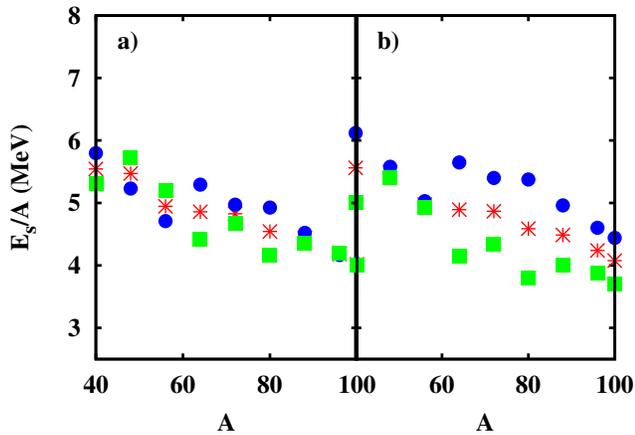}
\caption{
(Color online)
Hartree-Fock calculations. 
Surface energy per nucleon (red stars) and its
local (blue circles) and non-local (green squares) components multiplied by $2$,
for symmetric nuclei, as a function of the mass number.
Left (right) panel: Coulomb energy excluded (included).
}
\label{Fig_sym_TheoMagik}
\end{figure}

From Eq.~(\ref{eq_theo_magik_3D}), it is clear that the degree of violation of equality (\ref{magic}) will
depend on the functional, as well as on the variational model.
This  point is  illustrated in
Figure~\ref{Fig_sym_TheoMagik}, which  shows again the decomposition of the surface energy $E_s$
into local (blue circles) and non-local parts (green squares), calculated numerically
from spherical Hartree-Fock calculations.
In the calculations presented 
in the left panel the Coulomb energy, which breaks the equality $E_s^L=E_s^{NL}$ even in one-dimensional matter~\cite{thesis_aymard}, is artificially switched off. We can see that the 
Euler-Lagrange result in the slab geometry Eq.~(\ref{magic}) is reasonably well verified within 10\%, especially for medium-heavy nuclei $A\gtrsim 90$. 
This shows that the approximate equality between local and non-local terms is not limited to the ETF 
variational principle, but it is also verified by the Hartree-Fock variational solution.  
However, if the Coulomb interaction is included (right panel), the self-consistent modification of the Hartree-Fock density profile due to Coulomb is sufficient to lead to a strong violation of the equality between local and non-local terms, going up to 50\%.
 
This discussion shows that the exact shape of the density profile, and in particular the exact value of the diffuseness parameter, are  not important for the determination of the global surface energy,
but are crucial for a correct separation of local and non-local components. In practice it is very difficult to extract precisely the diffusivity coefficient  from theory or experiment: as we have seen in Fig.~\ref{Fig_sym_diffuseness_isoscal}, the diffuseness extracted from the Hartree-Fock variational density profile is very different from the ETF value, though the energies are close. Moreover the equality theorem is violated both because of curvature effects and of isospin symmetry breaking terms 
which cannot be neglected even in symmetric nuclei because of the Coulomb interaction. For all these reasons,  we conclude that the contribution from non-local terms cannot be estimated from the local part making use of  Eq.~(\ref{magic}). As a consequence, nuclear surface properties cannot be understood without mastering the gradient and spin-orbit terms of the energy functional.

\section{Asymmetric nuclei}
\label{sec_asym_nucl_vacuum}

We now turn to examine the general problem of an ETF analytical  mass model for 
asymmetric nuclei, which requires the introduction of the proton and neutron
density profiles as two independent degrees of freedom.
In this general case, the ETF energy integral cannot be evaluated analytically.
The usual approach in the literature consists in calculating the integral numerically, with density profiles which 
are either parametrized~\cite{chamel_etf,mekjian,pana}, or determined with a variational calculation~\cite{Centelles1990,centelles1,centelles2,agrawal}.
The limitation of such approaches is that the decomposition of the total binding energy into its different components
(isoscalar, isovector, surface, curvature, etc.) out of a numerical calculation is not unambiguous nor unique~\cite{aymard}.
Moreover, a numerical calculation makes it hard to discriminate the specific influence of the different physical parameters (EOS properties, finite range, spin orbit, etc)
on quantities like the surface symmetry energy or the neutron skin.

As a consequence, correlations between observables and physical parameters requires a statistical analysis based on a large set of very different models. In this way, one may hope that the obtained correlation is not spuriously induced by the specific form of the effective interaction~\cite{ducoin2010}. 
The correlation may also depend on several physical parameters and the statistical analysis becomes quite complex~\cite{margueron}.

Earlier approaches in the literature have introduced approximations in order to keep an analytical evaluation possible~\cite{krivine_iso}.
These approximations however typically neglect the presence of a neutron skin, and more generally of inhomogeneities
in the isospin distribution~\cite{mekjian}. As a consequence, the results are simple and transparent, but their validity out of the 
stability valley should be questioned.

One of the main applications of the present work concerns the production of reliable mass tables for an extensive use in astrophysical applications~\cite{compose}.
For this reason, we aim at expressions which stay valid approaching the driplines.
In the specific application to the neutron star inner crust, even more exotic nuclei far beyond the driplines are known to be populated~\cite{baym,negele}.
We will not consider this situation in the present paper, because a correct treatment of nuclei beyond the dripline imposes considering 
the presence of both bound and unbound states 
which modify the density profiles and leads to the emergence of a nucleon gas. 
Optimal parametrized density profiles
have been proposed for this problem~\cite{aymard,pana,esym}, but the developement of systematic approximations to analytically integrate the ETF functional in the presence of a gas is a delicate issue, which will be addressed in a forthcoming paper \cite{thesis_aymard}.

%
%
\subsection{Decomposition of the nuclear energy}
\label{sec_asym_changing_sym}

The presence of two separate good particle quantum numbers, $N$ and $Z$, implies 
that we have to work with a $2$-dimensional problem, and introduce,
in addition to the total density profile Eq.~(\ref{eq_sym_density_profile}), an additional
degree of freedom. 
Concerning the energy functional, 
it is customary to split it into an isoscalar and an isovector component:
\be
\mathcal{H}[\rho,\rho_3]=\mathcal{H}^{IS}[\rho,\rho_3=0]+\mathcal{H}^{IV}[\rho,\rho_3] 
\label{eq_energy_is_iv}
\ee
with:
\bea
\mathcal{H}^{IS}[\rho,\rho_3]       & = &\frac{\hbar^2}{2m}\tau  + C_{eff}\rho\tau + ( C_0 +C_3\rho^{\alpha})\rho^2 \nonumber \\
&+& C_{fin}(\boldsymbol{\nabla}\rho)^2 
+C_{so}\mathbf{J}\cdot\boldsymbol{\nabla}\rho \label{eq_energy_Skyrme_is} , \\
\mathcal{H}^{IV}[\rho,\rho_3]       & = &  \mathcal{H}^{IS}[\rho,\rho_3] - \mathcal{H}^{IS}[\rho,\rho_3=0] \nonumber \\
&+& D_{eff}\rho_3\tau_3 + (D_0  + D_3\rho^{\alpha})\rho_3^2 \nonumber \\
&+& D_{fin}(\boldsymbol{\nabla}\rho_3)^2   + D_{so}\mathbf{J}_3\cdot\boldsymbol{\nabla}\rho_3 
,
\label{eq_energy_Skyrme_iv}
\eea
where we have introduced the local isoscalar and isovector particle densities, kinetic densities and spin-orbit density vectors. 
Isoscalar densities are given by the sum of the corresponding neutron and proton densities, while isovector densities (noted with the subscript "3")
are given by their difference. 
As for symmetric matter, the semi-classical Wigner-Kirkwood development in $\hbar$ allows expressing all these 
densities in terms of the local isoscalar $\rho=\rho_n+\rho_p$ and isovector $\rho_3=\rho_n-\rho_p$ density profiles, as well as their gradients. 
In equation~(\ref{eq_energy_Skyrme_is}), the isoscalar energy density  also depends on $\rho_3$ because of 
the presence of the kinetic densities $\tau=\tau_n+\tau_p$ 
which cannot be written as a function of $\rho$ only.
Therefore, to truly obtain the isoscalar part in Eq.~(\ref{eq_energy_is_iv}), we have to consider $\mathcal{H}^{IS}[\rho,\rho_3=0]$.
The iso-vector energy density Eq.~(\ref{eq_energy_Skyrme_iv}) contains therefore terms which explicitly depend 
on the isovector densities, but also an isovector contribution of the so-called isoscalar component $\mathcal{H}^{IS}$.
Detailed expressions, and definition of parameters are given in appendix~\ref{sec_effective_interaction}.
\subsubsection{Isospin inhomogeneities}
\label{sec_asym_density_profile}
Concerning the density profiles,
we choose to work with the total density $\rho(r)$ and with the
proton density profile $\rho_p(r)$.
Alternatively, we could as well have used $(\rho,\rho_n)$ or $(\rho_p,\rho_n)$ as independent variables, and  we have 
checked that these different representations lead to the same level of reproduction of full Hartree-Fock calculations.
The total density is parametrized by Eq.~(\ref{eq_sym_density_profile}), where now the saturation density parameter $\rho_{sat}$ corresponds 
to the equilibrium density reached in asymmetric matter~\cite{pana}.
This density
depends on the asymmetry $\delta$ which
represents the nucleus bulk asymmetry, defined below:
\bea
\rho_{sat}(\delta) = \rho_{sat}(0) \left( 1 - \frac{3 L_{sym}\delta^2}{K_{sat}+K_{sym}\delta^2} \right).
\label{eq_asym_rho0}
\eea
In this expression,
$K_{sat}=9\rho_{sat}^2\partial^2(\mathcal{H}/\rho)/\partial\rho^2|_{\rho_{sat}}$ is the nuclear (symmetric) matter incompressibility,
and
$L_{sym}=3\rho_{sat}\partial (\mathcal{H}_{sym}/\rho)/\partial \rho |_{\rho_{sat}}$  and 
$K_{sym}=9\rho_{sat}^2\partial^2 (\mathcal{H}_{sym}/\rho)/\partial \rho^2 |_{\rho_{sat}}$ 
are the slope and curvature of the symmetry energy at (symmetric) saturation, where 
we have introduced the usual definition of the symmetry energy density :
\bea
\mathcal{H}_{sym}= \frac{1}{2}\rho^{2} \left.\frac{\partial^2 \mathcal{H}}{\partial \rho_3^{2}}\right|_{\rho_3=0}.
\eea
%
As a consequence, the radius parameter $R$ entering Eq.~(\ref{eq_sym_density_profile})
also depends on the nucleus bulk asymmetry $\delta$.
Indeed, in Eq.~(\ref{eq_sym_3D_R_1}), the equivalent homogeneous sphere radius now reads
$R_{HS}=A^{1/3} r_{sat}(\delta)$, where the mean  radius per nucleon is $r_{sat}(\delta)=\left( \frac{4}{3}\pi\rho_{sat}(\delta) \right)^{-1/3}$.

The proton density profile is parametrized as an independent 
Fermi function~\cite{pana}:
\bea
\rho_p(r) = \rho_{sat,p} F_p(r)
\;\; ; \;\;
F_p(r) = \left( 1+\e^{ (r-R_p)/a_p } \right)^{-1}.
\label{eq_asym_proton_density_profile}
\eea
In equation~(\ref{eq_asym_proton_density_profile}),
the proton radius parameter $R_p$  is determined, similarly to  Eq.~(\ref{eq_sym_3D_R_1}),
by proton number conservation as:
\bea
R_p =
R_{HSp} 
\left[
		1 - \frac{\pi^2}{3} 	\left( \frac{a_p}{R_{HSp}} \right)^2 
			+ O\left( \left(\frac{a_p}{R_{HSp}}\right)^4  \right)
\right]
\label{eq_asym_3D_Rp_1}
,
\eea
with $R_{HSp}(\delta) = Z^{1/3} r_{sat,p}(\delta)$ the equivalent homogeneous proton sphere radius,
$r_{sat,p}(\delta) = \left( \frac{4}{3}\pi\rho_{sat,p}(\delta) \right)^{-1/3}$, 
and where we assumed $a_p \ll R_p$ .

The diffusenesses $a$ and $a_p$ will be calculated in section~\ref{sec_asym_Gauss_apprx} 
by a minimization of the  surface energy,
as it has been done for symmetric nuclei in section~\ref{sec_sym_diffuseness} where $a_p=a$.
We can anticipate  that the isoscalar diffuseness $a$ will be modified with respect to the result of symmetric nuclei 
Eqs.~(\ref{eq_sym_diffuseness_SLAB}) and (\ref{eq_sym_diffuseness_3D}).


In order to have the correct bulk limit of infinite asymmetric
matter, the parameters $\rho_{sat}$ and $\rho_{sat,p}$ introduced 
in Eqs.~(\ref{eq_sym_density_profile}) and (\ref{eq_asym_proton_density_profile})
respectively represent the saturation densities of baryon and proton of asymmetric matter.
These densities are related to the properties of the Skyrme functional
and to the bulk asymmetry $\delta = 1 - 2 \rho_{sat,p}/\rho_{sat}$ by Eq.~(\ref{eq_asym_rho0}).

The bulk asymmetry differs from the global asymmetry 
$I = 1 - 2Z/A$ because of the competing effect of the Coulomb interaction and symmetry energy, which act
in opposite directions in determining the difference between the proton and neutron radii~\cite{centelles1,centelles2,ldm}:
\bea
\delta = \frac{  I + \frac{3a_c}{8Q} \frac{Z^2}{A^{5/3}}  }{  1+ \frac{9 J_{sym}}{4Q} \frac{1}{A^{1/3}}  }
.
\label{eq_asym_deltar}
\eea
In this equation, $J_{sym}=\mathcal{H}_{sym}[\rho_{sat}]/\rho_{sat}$ is the symmetry energy per nucleon at the saturation
density of symmetric matter,  
$Q$ is the surface stiffness coefficient, and $a_c$ is the Coulomb parameter.
Because of the  complex interplay between Coulomb and  skin effects, the bulk asymmetry $\delta$ of a globally symmetric $I=0$ nucleus is not zero, 
though small for nuclei in the nuclear chart. 
We have shown in Ref.~\cite{aymard} that accounting for the $\delta$ dependent saturation density 
gives a reasonably good approximation of the isospin symmetry breaking effects in $I=0$ nuclei.
A complete discussion on this point can be found in Ref.~\cite{dan03}. 

\begin{figure}[htbp]
\includegraphics[angle=270,width=\linewidth, clip]{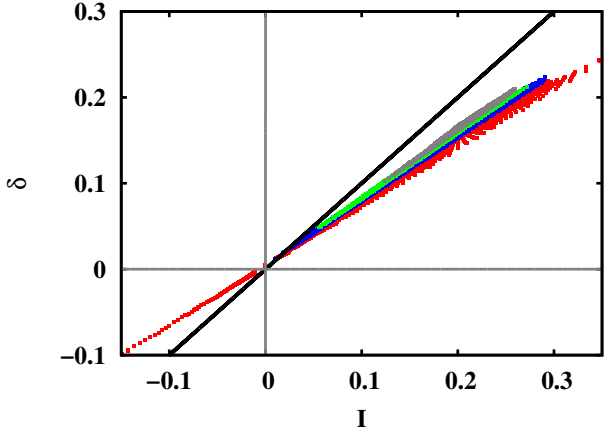}
\caption{ 
(Color online)
Bulk asymmetry Eq.~(\ref{eq_asym_rho0}) as a function of the global asymmetry $I$ 
for nuclei within the theoretical driplines evaluated from the SLy4 energy functional.
The different colors correspond to different intervals in mass number: 
$40 \leq A<100$ in red, 
$100\leq A<150$ in blue,
$150\leq A<200$ in green,
$A\geq 200$ in grey.
The function $y=x$ is also plotted (black).
}
\label{Fig:delta_I}
\end{figure}

As a consequence, the interval of $\delta$ is slightly smaller than the interval of $I$ over the periodic table.
The relation between the global asymmetry and the asymmetry in the nuclear bulk is shown in Fig.~\ref{Fig:delta_I}.
From this figure we can see that $\delta$ is a slowly increasing function of the global asymmetry $I$.
This value increases to $-0.1<\delta<0.3$ if we consider the ensemble of the heavy and medium-heavy 
nuclei within the driplines 
\cite{footnote2}.
It is also observed from Fig.~\ref{Fig:delta_I} that as the mass $A$ increases, $\delta$ becomes closer to $I$, as expected from the analytical
expression~(\ref{eq_asym_deltar}).

\begin{figure}[htbp]
\includegraphics[angle=270,width=\linewidth, clip]{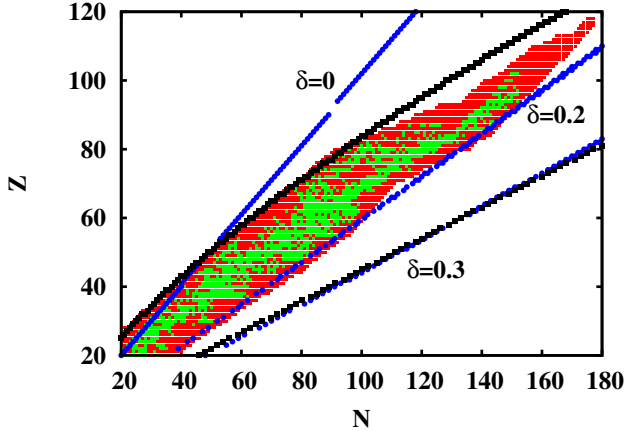}
\caption{ 
(Color online)
$\beta$-stable nuclei (green), unstable nuclei synthetized in the laboratory~\cite{nudat} (red),
theoretical neutron and proton driplines evaluated from the SLy4 energy functional (black squares) and 
some iso-$\delta$ lines (blue dots) are plotted in the $N,Z$ plane. 
}
\label{Fig:limits_delta}
\end{figure}

Figure~\ref{Fig:limits_delta} shows in the $(N,Z)$ plane
the heavy and medium-heavy measured nuclei, the theoretical neutron and proton driplines evaluated from the SLy4 energy functional, and some iso-$\delta$ lines.
We can see that all $A\lesssim 40$-isotopes ever synthesized in the laboratory lay between $\delta\approx 0$ and $\delta\approx 0.2$.
Furthermore, the theoretical neutron dripline well
matches with the iso-$\delta$ line $\delta\approx 0.3$, which roughly corresponds to $I \approx 0.4$.

This means that in the following, we will be interested in approximations producing reliable formulae up to $\delta\approx 0.3$.
\subsubsection{Bulk energy: limit of asymmetric nuclear matter}
\label{sec_asym_bulk_energy}

Following the same procedure as for the symmetric case, we can define the bulk energy
in asymmetric matter as:
\bea
E_b(\delta) &=& \mathcal{H}_{sat}(\delta) V_{HS}(\delta) = \lambda_{sat}(\delta) A,
\label{eq_asym_bulk}
\\
\mathcal{H}_{sat}(\delta) &=& \lambda_{sat}(\delta) \rho_{sat}(\delta)
,
\label{eq_asym_bulk_dens}
\eea
where $V_{HS}(\delta)=4/3\pi R_{HS}^3(\delta)=A/\rho_{sat}(\delta)$ is the equivalent homogeneous sphere volume and
$\lambda_{sat}(\delta)$ corresponds to the chemical potential of asymmetric nuclear matter:
\begin{align}
\left. \frac{\partial \mathcal{H}}{\partial \rho} \right|_{[\rho_{sat}(\delta),\rho_{sat,3}(\delta)]}
=
\lambda_{sat}(\delta)
=
\left. \frac{\mathcal{H}}{\rho} \right|_{[\rho_{sat}(\delta),\rho_{sat,3}(\delta)]}
.
\label{eq_lambda_asym_nm}
\end{align}
Here, $\rho_{sat,3} = \rho_{sat} - 2\rho_{sat,p}$, and 
the total energy density $\mathcal{H}$ is given by Eq.~(\ref{eq_energy_is_iv}).
\subsubsection{Decomposition of the   surface energy}
\label{sec_asym_non_bulk_energy}
The surface energy $E_{s}(\delta)$ corresponds to finite-size effects and can be decomposed, as in the symmetric case in section~\ref{sec_sym_nucl_vacuum}, 
into the plane surface, the curvature, and the higher order terms. 
It is defined as the difference between the total and the bulk $E_b(\delta)$ energy,
\begin{align}
E_{s}(\delta) & =
\int \mathrm d\mathrm{\mathbf{r}} \mathcal{H}[\rho,\rho_3]
- \mathcal{H}_{sat}(\delta) V_{HS}(\delta)
\nonumber \\ & =
4\pi \int_0^{\infty} \mathrm dr \big\{  \mathcal{H}[\rho,\rho_3] - \lambda_{sat}(\delta) \rho \big\} r^2
.
\label{eq_asym_3D_def_Enb}
\end{align}
Because of the isospin asymmetry, the Skyrme functional $\mathcal{H}$
now depends on the two densities $\rho$ and $\rho_3=\rho-2\rho_p$ 
and on the two gradients $\nabla\rho$ and $\nabla\rho_3=\nabla\rho-2\nabla\rho_p$.

Making again the decomposition of the energy density into an isoscalar  
(only depending on the total density)
and an isovector  component (depending on $\rho$ and $\rho_3$), we get from Eqs.~(\ref{eq_energy_Skyrme_is}) and~(\ref{eq_energy_Skyrme_iv}):
\bea
E_{s} = E^{IS}_{s} + E^{IV}_{s}
,
\label{eq_energy_decompo_sym_asym}
\eea
with
\begin{align}
E^{IS}_{s}   & = 
4\pi \int_0^{\infty} \mathrm dr \left\{  \mathcal{H}^{IS}[\rho,\rho_3=0] 
- \frac{\mathcal{H}^{IS}[\rho_{sat},\rho_{sat,3}=0]}{\rho_{sat}(\delta)} \rho  \right\} r^2
\nonumber \\
& = 
4\pi \int_0^{\infty} \mathrm dr \left\{  \mathcal{H}[\rho,\rho_3=0] 
- \frac{\mathcal{H}[\rho_{sat},\rho_{sat,3}=0]}{\rho_{sat}(\delta)} \rho  \right\} r^2
\label{eq_energy_decompo_sym} ,
\\
E^{IV}_{s}  & =
4\pi \int_0^{\infty} \mathrm dr \left\{  \mathcal{H}^{IV}[\rho,\rho_3] 
- \frac{\mathcal{H}^{IV}[\rho_{sat},\rho_{sat,3}]}{\rho_{sat}(\delta)} \rho  \right\} r^2
\nonumber \\
& = 
4\pi \int_0^{\infty} \mathrm dr \left\{  \mathcal{H}[\rho,\rho_3] 
- \frac{\mathcal{H}[\rho_{sat},\rho_{sat,3}]}{\rho_{sat}(\delta)} \rho  \right\} r^2
- E^{IS}_{s}
.
\label{eq_energy_decompo_asym}
\end{align}
It is interesting to remark that Eq.~(\ref{eq_asym_3D_def_Enb}) is not the only possible definition of the surface 
energy in a multi-component system.
Indeed in a two-component system, 
there are two possible definitions of the surface energy which depend on the definition of
the bulk energy in the cluster~\cite{Myers1985,Farine1986,Centelles1998}:
the first one is given by Eq.~(\ref{eq_asym_3D_def_Enb}) and corresponds to identifying the bulk energy 
of a system of $N$ neutrons and $Z$ protons to the energy of an equivalent
piece of nuclear matter . 
The second definition $E_{s}\equiv E-\mu_n N-\mu_p Z +pV$ corresponds to the grandcanonical thermodynamical Gibbs definition,
and gives the quantity to be minimized in the variational calculation 
 conserving proton and neutron number. Though this second definition has been often employed in the ETF literature~\cite{Myers1985,Farine1986,Centelles1998,agrawal}, the first one Eq.~(\ref{eq_asym_3D_def_Enb})
is the most natural definition in the present context. Indeed, using the decomposition Eq.~(\ref{eq_energy_is_iv}) between isoscalar and isovector 
energy densities, only this definition allows recovering for the isoscalar energy, the  results of section \ref{sec_sym_nucl_vacuum} concerning symmetric matter.
Moreover, we have shown in Ref.~\cite{aymard} that the best reproduction of full Hartree-Fock calculations is achieved considering
that the bulk energy in a finite nucleus scales with the bulk asymmetry $\delta$ as in Eq.~(\ref{eq_asym_3D_def_Enb}), rather than with the total asymmetry $I$, as it is implied
by the Gibbs definition. 

Let us first concentrate on the isoscalar surface energy.
The dependence of the surface energy on the bulk asymmetry $\delta$ implies that its decomposition
into an isoscalar and an isovector part is not straightforward. Indeed,
although the isoscalar energy $E^{IS}_{s}$ does not depend on the isospin asymmetry profile $\rho_3(r)$,
it does depend on the bulk isospin asymmetry $\delta$ through the isospin dependence of the saturation density $\rho_{sat}(\delta)$
 appearing in the density profile $\rho$ Eq.~(\ref{eq_sym_density_profile}).
Moreover, in Eq.~(\ref{eq_energy_decompo_sym}) the isoscalar bulk term which is removed depends directly on $\delta$ too,
because of the equivalent volume $V_{HS}=A/\rho_{sat}(\delta)$.
The quantity $E_{s}^{IS}$ has therefore an implicit dependence  on isospin asymmetry $\delta$.

The isoscalar surface energy can be calculated exactly for any nucleus of any asymmetry, with the expressions
developed in section~\ref{sec_sym_nucl_vacuum}. In particular we can distinguish a plane surface, a curvature, and a mass 
independent term:
\bea
E_{s}^{IS} = E_{surf}^{IS} + E_{curv}^{IS} + E_{ind}^{IS} 
+ O\left( \left(\frac{a(A,\delta)}{r_{sat}(\delta)}\right)^4 A^{-1/3} \right)
,
\label{eq_asym_Enbsym}
\eea
with an identical result as in Eqs.~(\ref{eq_sym_3D_Es_cl_surf}), (\ref{eq_sym_3D_Es_cl_curv}), (\ref{eq_sym_3D_Es_cl_ind}), namely:
\begin{align}
E_{surf}^{IS} &
= \left[ \mathcal{C}^{L}_{surf} 
+ \frac{1}{a^2(A,\delta)} \mathcal{C}^{NL}_{surf}  \right] \frac{a(A,\delta)}{r_{sat}(\delta)} A^{2/3} 
\label{eq_asym_3D_Es_cl_surf} ,
\\
E_{curv}^{IS} &
= \left[ \mathcal{C}^{L}_{curv} 
+ \frac{1}{a^2(A,\delta)} \mathcal{C}^{NL}_{curv}\  \right] \left( \frac{a(A,\delta)}{r_{sat}(\delta)} \right)^2 A^{1/3} 
\label{eq_asym_3D_Es_cl_curv} ,
\\
E_{ind}^{IS} & 
= \left[ \mathcal{C}^{L}_{ind} 
+  \frac{1}{a^2(A,\delta)} \mathcal{C}^{NL}_{ind}  \right]  \left( \frac{a(A,\delta)}{r_{sat}(\delta)} \right)^3
\label{eq_asym_3D_Es_cl_ind}
.
\end{align}
The local $\mathcal{C}^L_{i}$ and non-local $\mathcal{C}^{NL}_{i}$ functions are given 
by Eqs.~(\ref{eq_sym_coeff_L}) and (\ref{eq_sym_coeff_NL}), where the saturation density now depends on asymmetry 
 $\rho_{sat}=\rho_{sat}(\delta)$ through Eq.~(\ref{eq_asym_rho0}).
The other difference with respect to the case of symmetric nuclei 
Eqs.~(\ref{eq_sym_3D_Es_cl_surf}), (\ref{eq_sym_3D_Es_cl_curv}), (\ref{eq_sym_3D_Es_cl_ind}), 
is that now the diffuseness depends on the asymmetry $\delta$.

\begin{figure}[htbp]
\includegraphics[angle=270,width=\linewidth, clip]{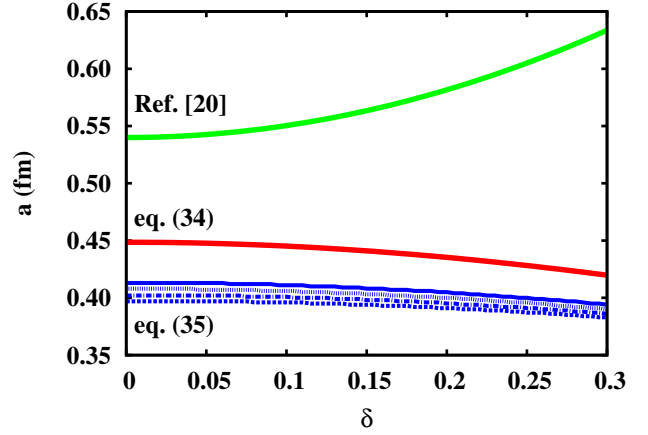}
\caption{
(Color online)
Diffuseness as a function of the isospin asymmetry,
for four isobaric  chains ($A=400$: full lines, $A=200$: dotted lines, $A=100$: dashed-dotted lines, $A=50$: dashed lines).
Red lines: calculations using the slab diffuseness Eq.~(\ref{eq_sym_diffuseness_SLAB}).
Blue lines: calculations using the spherical diffuseness Eq.~(\ref{eq_sym_diffuseness_3D}).
Green lines: calculations using the quadratic diffuseness, fitted from HF density profiles in Ref.~\cite{pana}.
}
\label{Fig_sym_diffuseness_isoscal_deltar}
\end{figure}

Since the analytical expressions of the isoscalar surface energy $E_s^{IS}$ are the same as in
symmetric nuclei, the same accuracy and conclusions as in section~\ref{sec_sym_nucl_vacuum} are dressed:
we can variationally evaluate the isoscalar diffuseness $a$, solving equation~(\ref{eq_sym_diffuseness_3D}), 
or using equation~(\ref{eq_sym_diffuseness_SLAB})  which amounts to neglecting terms  varying slower than $A^{2/3}$.
Though we have considered only isoscalar terms, the diffuseness $a$ does depend 
on the isospin asymmetry $\delta$ because of the $\delta$ dependence of the saturation density.
These results, as well as the fit from HF density profiles~\cite{pana}, 
where mass independence and quadratic behaviour in $\delta$ is assumed (that is: $a=C_1+C_2\delta^2$), 
are shown in Fig.~\ref{Fig_sym_diffuseness_isoscal_deltar}.
Concerning the mass-dependence of Eq.~(\ref{eq_sym_diffuseness_3D}) (blue lines labelled "Eq.~(\ref{eq_sym_diffuseness_3D})"),
we observe a slight spread for masses from $A=50$ to $A=400$, 
corroborating both the mass independence assumption in the HF fit~\cite{pana} 
and the previous conclusions in section \ref{sec_sym_diffuseness}: to obtain the diffuseness
we can neglect the mass dependence and limit to terms $\propto A^{2/3}$ (red line, labelled "Eq.~(\ref{eq_sym_diffuseness_SLAB})").
However, one can see that the dependence found from the variational equation is opposite to the one
exhibited by the fit to HF results:
the diffuseness decreases with $\delta$ instead of increasing.
It is difficult to believe that such a huge and qualitative difference might come from the difference between ETF and HF.
The discrepancy rather suggests  that the variational procedure should include the isovector energy  to obtain 
the correct behaviour of the diffuseness with the isospin asymmetry.
Indeed, we will see in section~\ref{sec_asym_Gauss_apprx} that adding the isovector part reverses the trend.

This discussion shows that, in the case of asymmetric nuclei, Eq.~(\ref{eq_sym_diffuseness_SLAB}) which only takes into account the isoscalar terms,
is not a good approximation to find the diffuseness.
\begin{figure}[htbp]
\includegraphics[angle=270,width=\linewidth, clip]{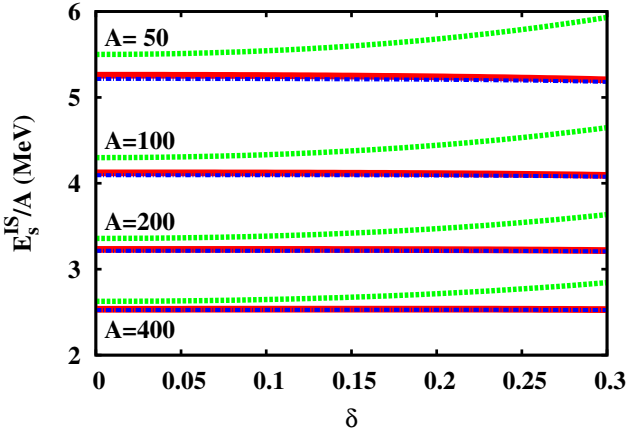}
\caption{
(Color online)
Isoscalar surface energy per nucleon as a function of the isospin asymmetry,
for four isobaric chains.
Full red lines: calculations using the slab diffuseness Eq.~(\ref{eq_sym_diffuseness_SLAB}).
Dash-dotted blue lines: calculations using the spherical diffuseness Eq.~(\ref{eq_sym_diffuseness_3D}).
Dashed green lines: calculations using the quadratic diffuseness, fitted from HF density profiles in Ref.~\cite{pana}.
}
\label{Fig_sym_diffuseness_isoscal_deltar_E}
\end{figure}
This statement is confirmed by Fig.~\ref{Fig_sym_diffuseness_isoscal_deltar_E},
where the isoscalar surface energy per nucleon is plotted for different isobaric chains and for different prescriptions for the diffuseness.
The full red and the dashed-dotted lines stand for the diffuseness given by Eq.~(\ref{eq_sym_diffuseness_SLAB}) and 
Eq.~(\ref{eq_sym_diffuseness_3D}) respectively. There is almost no difference in the isoscalar surface energy for these
two prescriptions. In addition, the observed $\delta$ dependence  is extremely weak.
The isoscalar surface energy evaluated with the quadratic diffuseness~\cite{pana} is represented in dashed green line. 
A qualitative and quantitative difference is observed with respect to the two other curves.
This indicates again that the isoscalar and isovector component of the surface energy cannot be treated separately, and the
correct $\delta$ dependence of the isoscalar surface energy, as well as of the isoscalar diffuseness, requires to consider
the total surface energy in the variational principle.

 It is also interesting to analyse the $\delta$ dependence of the surface symmetry energy based on the fitted quadratic diffuseness: 
its sign is positive, which contrasts with studies based on liquid-drop parametrizations of the 
nuclear mass~\cite{dan,ldm,frdm,reinhardt,pei,douchin}. This behavior is due to our choice of definition
of the surface in a two component system, as discussed at length in Ref.~\cite{aymard}.

%
%
%
%
\subsection{Approximations for the isovector energy}
\label{sec_asym_Gauss_apprx}
In this section, we focus on the residual isovector surface part $E_{s}^{IV}$ defined by Eq.~(\ref{eq_energy_decompo_asym}),
which cannot be written as integrals of Fermi functions as in the previous sections.
Indeed, the isovector density $\rho_3$  appearing in the energy density is not a Fermi function,
meaning that it cannot be analytically integrated to evaluate  $E_{s}^{IV}$.
Approximations are needed to develop an analytic expression for this part of the energy,
and we will consider in the following two different approaches.
At the end, we will verify the accuracy of our final formulae, 
in comparing the analytical expressions with HF calculations.
\subsubsection{No skin approximation}
\label{sec_asym_TK}
As a first approximation, we  neglect all inhomogeneities in the isospin distribution in the same spirit as Ref.~\cite{krivine_iso}.
This simplification consists in replacing the isospin asymmetry profile $\rho_3(r)/\rho(r)$ in Eq.~(\ref{eq_energy_decompo_asym})
by its mean value $\langle\delta\rangle$.
 Within this approximation, the local isovector energy only depends on the total baryonic density profile $\rho$ defined Eq.~(\ref{eq_sym_density_profile}),
and the non-local isovector part, involving gradients $\nabla\rho_3$, is identically zero.
In other words, this approximation amounts neglecting the non-local contribution to the isovector surface energy. 

Integrating in space the equality $\rho_3(r) = \langle\delta\rangle \rho(r)$
we immediately obtain that the mean value of the isospin distribution
is given by the global asymmetry of the nucleus:
\bea
\langle\delta\rangle = \frac{N-Z}{A} = I.
\label{eq_noSkin_delta_I}
\eea

In particular, in this approximation, the bulk isospin asymmetry $\delta$ is equal to the global asymmetry $I$,
at variance with the more elaborated relation between $\delta$ and $I$ given by Eq.~(\ref{eq_asym_deltar}).
In neglecting isospin inhomogeneities, we indeed neglect both neutron skin and Coulomb effects
which are responsible for the difference between $\delta$ and $I$.
Consequently in this section, the saturation density $\rho_{sat}$ of asymmetric matter 
is still given by Eq.~(\ref{eq_asym_rho0}), but replacing $\delta$ by $I$.
This no-skin approximation therefore modifies the bulk energy Eq.~(\ref{eq_asym_bulk}),
and  the isoscalar energy Eq.~(\ref{eq_sym_3D_def_Enb}).
  
The choice of $I$ instead of $\delta$  to compute the saturation density
only slightly worsens the predictive power of the total ETF energy with respect to
Hartree-Fock calculations, but the relative weight between bulk and surface energies
is drastically modified. In particular, this change of variable switches the sign of the symmetry surface energy~\cite{aymard}.

The obvious advantage is that analytical results can be obtained without further approximations than the ones 
developed in section~\ref{sec_sym_energy}, as we now detail.

Replacing $\rho_3(r)$ by $I\rho(r)$ and $\rho_{sat,3}(\delta)$ by $I\rho_{sat}(I)$ in Eq.~(\ref{eq_energy_decompo_asym}),
allows to express the energy density as a function of $\rho(r)$ only.
Thus we can follow the same procedure as for symmetric nuclei in section~\ref{sec_sym_energy},
and analytically integrate the energy density.
Making a quadratic expansion in $I$ for the kinetic densities $\tau_3$ gives the following expressions:
\bea
E_s^{IV} & = &
\mathcal{C}^{IV}_{surf}\left(\rho_{sat}(I),\mathbf{X}_{sky}^{IV}\right)   \frac{a(A,I)}{r_{sat}(I)} A^{2/3} I^2
\nonumber \\ 
&+& \mathcal{C}^{IV}_{curv}\left(\rho_{sat}(I),\mathbf{X}_{sky}^{IV}\right)   \left( \frac{a(A,I)}{r_{sat}(I)} \right)^2 A^{1/3} I^2
\nonumber \\
&+& \mathcal{C}^{IV}_{ind} \left(\rho_{sat}(I),\mathbf{X}_{sky}^{IV}\right)   \left( \frac{a(A,I)}{r_{sat}(I)} \right)^3  I^2
\nonumber \\
&+&  o \left(  \left( \frac{a(A,I)}{r_{sat}(I)} \right)^4 A^{-1/3} I^2 \right),
\label{eq_noSkin_IV}
\eea
where $\mathbf{X}_{sky}^{IV}=\left\{C_{eff}, \alpha, D_{eff}\right\}$ 
stands for the effective interaction parameters appearing in the isovector local terms.
The coefficients $\mathcal{C}_{i}^{IV}$ are given by:
\begin{widetext}
\bea
\mathcal{C}^{IV}_{surf}  & = &
3
\left\{
	 C_{kin}
				\left[
					 \frac{5}{3} \eta^{(0)}_{5/3} \left( \frac{m}{3 m^*_{sat}}  + \Delta m_{sat,3} \right)
					 - \left( \frac{ \delta m_{sat} }{3} + \Delta m_{sat,3} \right)
				\right]
	- D_0 \rho_{sat}
	+ D_3 \rho_{sat}^{\alpha+1} \eta^{(0)}_{\alpha+2}
\right\} ,
\label{eq_sym_coeff_IV_surf}
 \\
\mathcal{C}^{IV}_{curv}  &  =&
6
\left\{
	 C_{kin}
				\left[
				 	  \frac{5}{3} \left( \eta^{(1)}_{5/3} - \frac{\pi^2}{6} \right) \left( \frac{m}{3 m^*_{sat}} + \Delta m_{sat,3} \right)
				 	 - \eta^{(0)}_{5/3} \left( \frac{\delta m_{sat}}{3} + \Delta m_{sat,3}  \right)	
				\right]	
+ D_3 \rho_{sat}^{\alpha+1} 
	\left(
		\eta^{(1)}_{\alpha+2}
		- \frac{\pi^2}{6}
	\right)
\right\}
\label{eq_sym_coeff_IV_curv} ,
 \\
\mathcal{C}^{IV}_{ind}  & =&
3
\left\{
	 C_{kin}
		    	\left[	
					\frac{5}{3}\left( \eta^{(2)}_{5/3} - \frac{2 \pi^2}{3} \eta^{(0)}_{5/3}  \right)
														\left( \frac{m}{3 m^*_{sat}} + \Delta m_{sat,3} \right)
					- \frac{2}{3}
					\left( 3\eta^{(1)}_{5/3} - \pi^2\right)  \left( \frac{\delta m_{sat}}{3} + \Delta m_{sat,3} \right)
			\right]
\right.\nonumber \\ && \left.  \vphantom{ \left( \frac{3\pi^2}{2} \right)^{2/3} }
	+ \frac{\pi^2}{3} D_0 \rho_{sat}
	+ D_3 \rho_{sat}^{\alpha+1} 
		\left(	
			\eta^{(2)}_{\alpha+2}
			- \frac{2 \pi^2}{3} \eta^{(0)}_{\alpha+2}
		\right)
\right\}
,
\label{eq_sym_coeff_IV} 
\eea
\end{widetext}
where $m/m^*_{sat} = ( m/m^*_{sat,n} + m/m^*_{sat,p} ) /2$,
$\delta m_{sat} = ( \delta m_{sat,n} + \delta m_{sat,p} ) /2$,
$\Delta m_{sat,3} = ( m/m^*_{sat,n} - m/m^*_{sat,p} ) /(2I) = ( \delta m_{sat,n} - \delta m_{sat,p} ) /(2I)$,
and where the coefficients $\eta_\gamma^{(k)}$ 
are defined by equation~(\ref{eq_appendix_gen_eta}).

As for the isoscalar energy, Eq.~(\ref{eq_noSkin_IV}) shows that the dominant finite-size effect is
a surface term ($\propto A^{2/3}$).
Additional finite-size terms, which would be absent in a slab configuration, are found in spherical nuclei.
As we have only considered the local part of the isovector energy,
we recover the same diffuseness dependence as in the local isoscalar terms Eqs.~(\ref{eq_sym_3D_Es_cl_L}) and~(\ref{eq_sym_3D_Es_cl_NL}).

We have seen in section~\ref{sec_sym_diffuseness} that the diffuseness $a$ can be obtained
by minimization of the energy per nucleon with respect to its free parameters.
In this no-skin approximation, the only non-constrained parameter of the model
is again the diffuseness parameter $a$, as for symmetric nuclei.
Therefore, we can apply Eq.~(\ref{eq_sym_diffuseness_min_E_a}) in order to obtain the ground state energy.
If we neglect the curvature and mass independent terms, 
we obtain an expression similar to Eq.~(\ref{eq_sym_diffuseness_SLAB}):
\begin{align}
a & = 
\sqrt{ 
		\frac{ \mathcal{C}^{NL}_{surf}(I)}{ \mathcal{C}^{L}_{surf}(I) + \mathcal{C}_{surf}^{IV}(I) I^2 } 
	}
,
\label{eq_asym_diffuseness_SLAB_noSkin}
\end{align}
where the coefficients $\mathcal{C}_{surf}^i$ depend on the saturation density $\rho_{sat}(I)$.
This expression corresponds to the diffuseness of one-dimensional semi-infinite asymmetric matter.
Considering all the terms of Eq.~(\ref{eq_noSkin_IV}), the diffuseness corresponding to the complete variational problem is 
given by the solution of
the following equation:
\bea
&3& \left( \mathcal C_{ind}^{L}   + \mathcal{C}_{ind}^{IV} I^2  \right)      \left( \frac{a}{r_{sat}} \right)^4
+
2  \left( \mathcal C_{curv}^{L}  + \mathcal{C}_{curv}^{IV} I^2 \right) A^{1/3}  \left( \frac{a}{r_{sat}} \right)^3 
\nonumber \\
&+&
\left( \left( \mathcal C_{surf}^{L} + \mathcal{C}_{curv}^{IV} I^2 \right)  A^{2/3} + \frac{1}{r_{sat}^2} \mathcal C_{ind}^{NL} \right)  \left( \frac{a}{r_{sat}} \right)^2 \label{eq_asym_diffuseness_3D_noSkin}  \\
&-&
\frac{1}{r_{sat}^2} \mathcal C_{surf}^{NL} A^{2/3}
= 0
.
\nonumber
\eea
\begin{figure}[htbp]
\centering
\includegraphics[angle=270,width=\linewidth, clip]{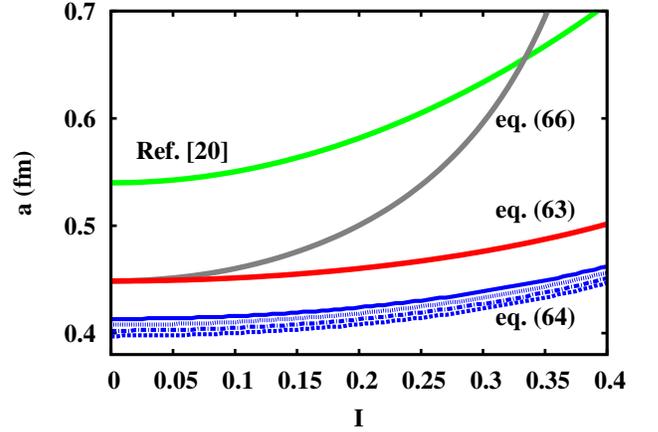}
\caption{
(Color online)
Diffuseness as a function of the global asymmetry,
for four isobaric  chains ($A=400$: full lines, $A=200$: dotted lines, $A=100$: dashed-dotted lines, $A=50$: dashed lines).
Red lines: calculations using the slab diffuseness Eq.~(\ref{eq_asym_diffuseness_SLAB_noSkin}).
Blue lines: calculations using the spherical diffuseness Eq.~(\ref{eq_asym_diffuseness_3D_noSkin}).
Green line: calculations using the quadratic diffuseness fitted from HF density profiles~\cite{pana}.
Grey line: calculations using the diffuseness Eq.~(\ref{eq_diffuseness_TK}), based on~\cite{krivine_iso}.
}
\label{Fig_apprx_TK}
\end{figure}

Figure~\ref{Fig_apprx_TK} displays the results of 
Eqs.~(\ref{eq_asym_diffuseness_SLAB_noSkin}) and~(\ref{eq_asym_diffuseness_3D_noSkin}).
At variance with Fig.~\ref{Fig_sym_diffuseness_isoscal_deltar} where we only took into account the isoscalar energy,
we can clearly see that adding the isovector energy to the variational procedure  leads to the expected behavior
of a diffuseness increasing with asymmetry.
 
This behavior shows the importance of the isovector part to correctly determine the isoscalar diffuseness $a$.
As for symmetric nuclei, we observe again that the mass dependence of the diffuseness calculated in the spherical case 
is negligible (only a slight spread of the blue curves, no spread in the red curves).

\begin{figure}[htbp]
\centering
\includegraphics[angle=270,width=\linewidth, clip]{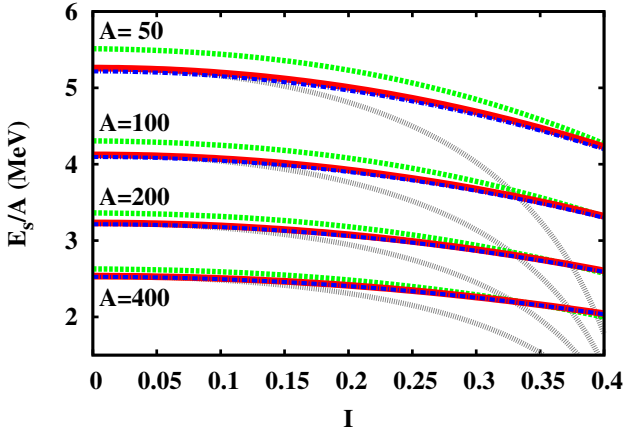}
\caption{
(Color online)
Total  surface energy per nucleon as a function of the global asymmetry,
for four isobaric  chains.
Full red lines: calculations using the slab diffuseness Eq.~(\ref{eq_asym_diffuseness_SLAB_noSkin}).
Dash-dotted blue lines: calculations using the spherical diffuseness Eq.~(\ref{eq_asym_diffuseness_3D_noSkin}).
Dashed green lines: calculations using the quadratic diffuseness fitted from HF density profiles~\cite{pana}.
Dotted grey lines: calculations using the diffuseness Eq.~(\ref{eq_diffuseness_TK}), based on~\cite{krivine_iso}.
}
\label{Fig_apprx_TK_E}
\end{figure}

The analytical total  surface energy $E_s=E_s^{IS}+E_s^{IV}$ per nucleon, 
given by Eqs.~(\ref{eq_sym_3D_Es_cl_L}), (\ref{eq_sym_3D_Es_cl_NL}) and~(\ref{eq_noSkin_IV}),
is plotted on Fig.~\ref{Fig_apprx_TK_E},
for different isobaric chains.
The results using the slab diffuseness (full red curves) 
are very close to the ones obtained by solving Eq.~(\ref{eq_asym_diffuseness_3D_noSkin}) (dash-dotted blue curves),
and to the ones using the numerical fit to HF calculations of Ref.~\cite{pana} (dashed green curves), even if 
the corresponding values for the $a$ parameter are very different.
The conclusions are thus the same as in section~\ref{sec_sym_diffuseness}:
although curvature (and mass independent) terms are important to reproduce the energetics,
they are not required to determine the diffuseness.
Therefore this latter can be well determined by the simplest expression, Eq.~(\ref{eq_asym_diffuseness_SLAB_noSkin}).

For completeness, we also compare our results to the approximation for the surface energy proposed in Ref.~\cite{krivine_iso}, 
and represented by grey curves in Figs.~\ref{Fig_apprx_TK} and \ref{Fig_apprx_TK_E}:
\bea
E_s & = &
E_s^{IS}(I=0) 
\nonumber \\
&+& 2 \left[   
	\frac{ E_s^{IS}(I=0) }{A^{2/3}} \frac{L_{sym}}{K_{sat}} -  \frac{ a\left( L_{sym} - \frac{K_{sym}}{12} \right)}{r_{sat}(I=0)}
\right]  A^{2/3}  I^2
.
\label{eq_formula_TK}
\eea
In Ref.~\cite{krivine_iso}, no expression for the diffuseness was proposed.
For consistency, we have determined the $a$ parameter entering Eq.~(\ref{eq_formula_TK}) 
by minimizing the surface energy given by the same equation, leading to:
\bea
a & = &
\sqrt{ 
		\frac{ 
			\mathcal{C}^{NL}_{surf}(I=0) \left( 1 + \frac{2L_{sym}}{K_{sat}} I^2 \right)
			}{ 
			\mathcal{C}^{L}_{surf}(I=0) \left( 1 + \frac{2L_{sym}}{K_{sat}} I^2 \right) - 2 \left( L_{sym} - \frac{K_{sym}}{12} \right) I^2 
			} 
	}
.
\label{eq_diffuseness_TK}
\eea
To obtain Eq.~(\ref{eq_formula_TK}) , the authors of Ref.~\cite{krivine_iso}  
 did the same approximation $\rho_3(r)=I\rho(r)$ as we made, neglected the curvature and constant terms,
and assumed the equality $E_s^{L}=E_s^{NL}$ for the isovector part
in order to evaluate the non-local isovector energy.
As we have shown in section~\ref{sec_sym_results}, this property fails in a three-dimensional system.
As a consequence, the diffuseness which is determined by the balance between local and non-local parts,
is overestimated (see Fig.~\ref{Fig_apprx_TK}) and finally leads to a largely underestimated energy, as seen in Fig.~\ref{Fig_apprx_TK_E}.

To check the accuracy of our analytical no-skin expression given by 
Eqs.~(\ref{eq_asym_Enbsym}), (\ref{eq_noSkin_IV}) and (\ref{eq_asym_diffuseness_SLAB_noSkin}),
we will quantitatively compare our analytical results 
with Hartree-Fock calculations
in section~\ref{sec_IV_compHF}.


%
\subsubsection{Gaussian approximation}
\label{sec_asym_gauss_tot}
To take into account isospin inhomogeneities, 
we develop in this section an alternative gaussian approximation to the isovector surface energy.
In particular, as in section~\ref{sec_asym_changing_sym}, 
we will distinguish the bulk asymmetry $\delta$ Eq.~(\ref{eq_asym_deltar}) from the global one $I$, 
which allows considering  skin and Coulomb effects.
This approximation is therefore expected to be more realistic than the no-skin procedure developed 
in section~\ref{sec_asym_TK}.
%
\begin{figure}[htbp]
\includegraphics[angle=270,width=\linewidth, clip]{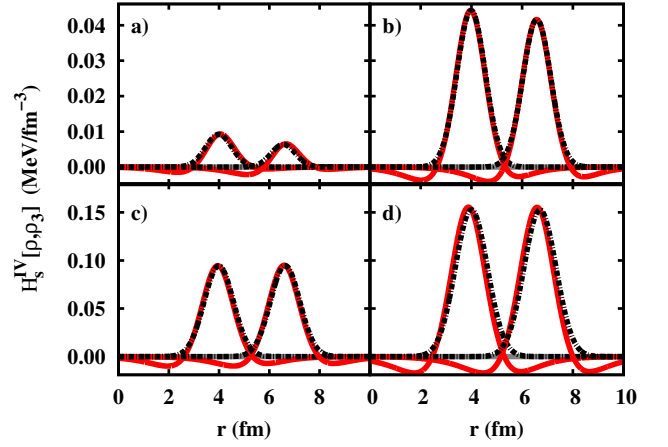}
\caption{ (Color online)
Numerical isovector energy density profile (red full lines) 
and  Gaussian approximation Eq.~(\ref{eq_asym_energy_density_tot_Gauss}) (black dashed-dotted lines)
for two masses $A=50$ (left curves of each panel) and $A=200$ (right curves of each panel).
a) $\delta=0.1$; b) $\delta=0.2$; c) $\delta=0.3$; d) $\delta=0.4$.
}
\label{Fig_sym_energy_profiles_Gauss_tot}
\end{figure}

Since  $E^{IV}_{s}$ is the surface isovector energy,
the corresponding energy density 
\bea
\mathcal{H}^{IV}_{s}[\rho,\rho_3] &=&
\mathcal{H}^{IV}[\rho,\rho_3]
-
\frac{  \mathcal{H}^{IV}[\rho_{sat},\rho_{sat3}]  }{  \rho_{sat}(\delta)  }  \rho
\label{eq_asym_energy_density_tot}
\eea
is negligible at the nucleus center, where $\rho\rightarrow\rho_{sat}$. This is shown in
Fig.~\ref{Fig_sym_energy_profiles_Gauss_tot}, which displays this quantity for several nuclei
in a representative calculation using the diffusenesses $a$ and $a_p$ from Ref.~\cite{pana},
and with the interaction SLy4.
Moreover, as it is a surface energy, the maximum is expected to be close to the surface radius $R$,
that is the inflection point where $\rho(R)=\rho_{sat}(\delta)/2$.
Thus we approximate the isovector energy density by a Gaussian peaked at $r=R$:
\bea
\mathcal{H}^{IV}_{s}(r) \simeq 
\mathcal{G}_{tot}(r) =
\mathcal{A}(A,\delta) \exp\left( - \frac{(r-R)^2}{2\sigma^2(A,\delta)} \right)
,
\label{eq_asym_energy_density_tot_Gauss}
\eea
where $\mathcal{A}$ is the maximum amplitude of the Gaussian and $\sigma^2$ its variance at $R$:
\bea
\mathcal{A}(A,\delta) &=& \mathcal{H}^{IV}_{s}[\rho(R),\rho_3(R)]
\label{eq_asym_ampl_tot} ,
\\
\sigma^2(A,\delta) &=& - \mathcal{A}(A,\delta)  \left( \frac{d^2 \mathcal{H}^{IV}_{s}}{\mathrm dr^2} \right)^{-1}_{r=R}
.
\label{eq_asym_var_tot}
\eea

Fig.~\ref{Fig_sym_energy_profiles_Gauss_tot} shows the quality of this Gaussian approximation
on the energy density profile for several nuclei. 
Each panel corresponds to a different  representative value of $\delta$:
$\delta=0.1$ (upper left) corresponds to most stable nuclei (see Fig.~\ref{Fig:limits_delta}); 
medium-heavy neutron rich nuclei synthesized in modern radioactive ion facilities lay around $\delta=0.2$ (upper right); 
the (largely unexplored) neutron drip-line closely corresponds to $\delta=0.3$ (lower left); 
the higher value $\delta=0.4$ (lower right)
is only obtained beyond the dripline, that is for nuclei which are in equilibrium with a neutron gas in the inner crust of neutron stars.

We can see that for all these very different asymmetries,
the exact energy density (full red lines) is indeed peaked at the equivalent hard sphere radius $R$.
However, we can notice that the profiles have small negative components.
We thus expect the  Gaussian approximation will  overestimate the isovector energy part.

As Gaussian functions and their moments are analytically integrable,
this approximation allows obtaining an analytical expression for the isovector energy 
$\displaystyle E_s^{IV}\approx 4\pi \int  r^2\mathcal{G}_{tot}(r)\mathrm d r$.
Indeed, neglecting the terms $\sim\e^{-R^2/(2\sigma^2)}$,
we obtain  (see appendix~\ref{sec_appendix_asym}):
\bea
E_{s}^{IV} & = &
2\left( 2\pi \right)^{3/2} \sigma(A,\delta, a, a_p) \mathcal{A}(A,\delta, a, a_p)
 r_{sat}^2(\delta) \nonumber \\
		&&\left[  
		A^{2/3} 
		+ \frac{\sigma^2(A,\delta, a, a_p)}{r_{sat}^2(\delta)}
		- \frac{2\pi^2}{3} \left( \frac{a(A,\delta)}{r_{sat}(\delta)} \right)^2
		\right]
,
\label{eq_asym_gauss_tot}
\eea
where we have highlighted the dependence on the nuclear mass number $A$, bulk asymmetry $\delta$, 
and diffusenesses $a(A,\delta)$ and $a_p(A,\delta)$ when they explicitly appear.
The neglected terms are of the order $(a/r_{sat})^4 A^{-2/3}$.
We can notice that the curvature term ($\propto A^{1/3}$) is missing. This is due to our approximation.
Indeed, we have assumed that the isovector energy  is symmetric with respect to the inflection point
for which the curvature is zero, such that the curvature is disregarded by construction.

Though equation (\ref{eq_asym_gauss_tot}) is an analytical expression, 
the explicit derivation of the amplitude $\mathcal{A}(A,\delta)$ and of the variance $\sigma(A,\delta)$
leads to formulae far from being transparent.
In particular, it is not clear how the different physical ingredients of the energy functional (compressibility, effective mass, symmetry energy) and of the nucleus properties (neutron skin, diffuseness) affect the isovector surface properties.
For this reason, we turn to develop  a further approximation for the isovector energy part $E^{IV}_{s}$ 
in terms of the nuclear matter coefficients $J$, $L$ and $K$, and of the neutron skin thickness.
Moreover, these approximations will allow to find a simple analytical expression for the diffuseness.

Making the usual quadratic assumption for the symmetry energy
$\mathcal{H}^{IV}[\rho,\rho_3] =  \mathcal{H}_{sym}[\rho] (\rho_3/\rho)^2$,
the amplitude $\mathcal{A}(A,\delta)$ Eq.~(\ref{eq_asym_ampl_tot}) reads
\bea
\mathcal{A}(A,\delta) = 
\mathcal{H}_{sym}[\rho(R)] \left( \frac{\rho_3(R)}{\rho(R)} \right)^2 
- 
\mathcal{H}_{sym}[\rho_{sat}(\delta)] \frac{\rho(R)}{\rho_{sat}(\delta)} \delta^2 .
\nonumber \\
\label{eq_apprx_energy_density_surf_IV}
\eea
In order to have a simpler explicit expression,
we make a density expansion of the symmetry energy per nucleon 
$e_{sym}[\rho]=\mathcal{H}_{sym}[\rho]/\rho$ around a density $\rho_*$, such that:
\bea
\mathcal{H}_{sym}[\rho] & = & 
\rho \left[
J_{*}
+ \frac{L_*}{3\rho_*} \left( \rho- \rho_* \right)
+\frac{K_*}{18\rho^2_*} \left( \rho- \rho_* \right)^2
\right]
,
\label{eq_exapnd_Hsym_rho}
\eea
where
$J_{*}=\mathcal{H}_{sym}[\rho_*]/\rho_*$,
$L_{*}=3\rho_{*}\partial (\mathcal{H}_{sym}/\rho_*)/\partial \rho |_{\rho_{*}}$  and 
$K_{*}=9\rho_{*}^2\partial^2 (\mathcal{H}_{sym}/\rho_*)/\partial \rho^2 |_{\rho_{*}}$.
As we can see in Eq.~(\ref{eq_apprx_energy_density_surf_IV}),  we need to evaluate the symmetry energy at two different densities:
 at $\rho_{sat}(\delta)$ and at the surface radius where $\rho(R)=\rho_{sat}(\delta)/2$.
For this reason, we will apply Eq.~(\ref{eq_exapnd_Hsym_rho}) to two different densities
$\rho_*=\rho_{sat}(0)$ and $\rho_*=\rho_{sat}(0)/2$.
At $\rho_*=\rho_{sat}(0)$, the coefficients $(J_*, L_*, K_*)$ are the usual symmetry energy coefficients $(J_{sym}, L_{sym}, K_{sym})$.
Their values for the Skyrme interaction SLy4 are 
$J_{sym}=32~\mathrm{MeV}$, $L_{sym}=46~\mathrm{MeV}$, and $K_{sym}=-119.8~\mathrm{MeV}$.  
At one half of the saturation of symmetric nuclear matter, $\rho_*=\rho_{sat}(0)/2$ we label the corresponding coefficients $(J_{1/2}, L_{1/2}, K_{1/2})$
which, for the Skyrme interaction SLy4, are 
$J_{1/2}=22.13~\mathrm{MeV}$, $L_{1/2}=38.6~\mathrm{MeV}$, and $K_{1/2}=-74~\mathrm{MeV}$.

Using the expansion around $\rho_*=\rho_{sat}(0)/2$ for the first term of Eq.~(\ref{eq_apprx_energy_density_surf_IV})
and around $\rho_*=\rho_{sat}(0)$ for the second one, we obtain, at second order in $\delta$:
\bea
\frac{\mathcal{A}(A,\delta)}{\rho_{sat}(0)} & = &
\frac{J_{1/2}}{8}  \left( \frac{\Delta R(a)}{a(A,\delta)} \right)^2 \nonumber \\
&+& \frac{J_{1/2}}{2} \left[
						\frac{\Delta R(a)}{a(A,\delta)} - \frac{1}{2} \left( \frac{\Delta R(a)}{a(A,\delta)} \right)^2
		 			\right] \delta
\nonumber \\
& + &  \frac{J_{1/2}}{2}  \Bigg[
						 \left( 1-\frac{J_{sym}}{J_{1/2}} \right)  - \frac{\Delta R(a)}{a(A,\delta)} 
\nonumber \\ 
& &  \hphantom{ \frac{J_{1/2}}{2} }  - 
					\frac{1}{4} \left( 1+ \frac{L_{sym}L_{1/2}}{J_{1/2}K_{sat}} \right) 
\left( \frac{\Delta R(a)}{a(A,\delta)} \right)^2
						\Bigg] \delta^2
.
\label{eq_analytical_amplitude}
\eea
Notice that the $K_{sym}$ parameter does not appear in this equation because of the truncation at second order in $\delta$.
In Eq.(\ref{eq_analytical_amplitude}), the isospin asymmetry inhomogeneities clearly appear through the quantity $\Delta R(a) = R(a) - R_p(a)$ 
which represents the neutron skin thickness:
\bea
\Delta R(a) = \Delta R_{HS} \left( 1 + \frac{\pi^2}{3} \frac{ a^2 }{ R_{HS} R_{HS,p}} \right)
,
\label{eq_neutron_skin_thickness}
\eea
where $\Delta R_{HS}(A,Z) = \Delta R(a=0,A,Z) = R_{HS}(A) - R_{HS,p}(Z)$ is 
the neutron skin thickness of nuclei theoretically described by hard spheres.
Moreover, we have considered the diffuseness difference $a-a_p$ as a second order correction with respect  to the neutron skin,
and have assumed $a=a_p$ in Eq.~(\ref{eq_analytical_amplitude}).
We have also used the following expansion in $\Delta R(a)/a$
to evaluate $\rho_3(R)$: 
\bea
2 \rho_p(R) &=& \rho_{sat,p}(\delta) \left[ 1 - \Delta R(a)/(2 a) \right] \\
&+& O\left( (\Delta R(a)/a)^3 \right). \nonumber 
\eea
Eq.(\ref{eq_analytical_amplitude}) gives a relatively simple and transparent expression of the isovector energy density at the nuclear surface, as a function of the EoS parameters. The situation is more complicated for 
the variance $\sigma(A,\delta)$ which also enters the isovector energy Eq.~(\ref{eq_asym_gauss_tot}). This quantity
involves the second spatial derivative of the energy density Eq.~(\ref{eq_asym_var_tot}), therefore
its explicit expression is not transparent, even with the previous simplifications.
Extra approximations are in order.

From Fig.~\ref{Fig_sym_energy_profiles_Gauss_tot}, we can observe that the width of the numerical gaussians, 
that is the values of $\sigma^2(A,\delta)$, is almost independent of the bulk isospin $\delta$.
This numerical evidence can be understood from the fact that the width gives a measure of the nucleus surface, which is 
mostly determined by isoscalar properties. It is therefore not surprising that the dominant isospin dependence
is given by the amplitude $\mathcal{A}$ which represents the isovector energy density at the surface.  
For this reason, we evaluate the variance at $\delta = 0$:
\bea
\sigma(A,\delta) & \approx & \sigma(A) =
\sqrt{   \frac{  2}  {1-\frac{K_{1/2}}{18 J_{1/2}} }   }   a_0
= \sigma_0
.
\label{eq_analytical_variance}
\eea
In this equation, $a_0$ stands for the diffuseness at $\delta=0$. We recall that this quantity does not depend on the nucleus mass if we do not take into account terms beyond surface in the variational approach discussed in section~\ref{sec_sym_diffuseness}. This approximate mass independence of the variance can be verified in Fig.~\ref{Fig_sym_energy_profiles_Gauss_tot}: 
the width of the two gaussians corresponding to $A=50$ and $A=200$ are very close.
Neglecting the isovector component at $\delta=0$,
the diffuseness is then given by the expression ~(\ref{eq_sym_diffuseness_SLAB}) valid for symmetric matter:
\bea
a_0 = \sqrt{ \mathcal{C}_{surf}^{NL}(\delta=0) / \mathcal{C}_{surf}^{L}(\delta=0) }. \label{a0}
\eea
Inserting Eqs.~(\ref{eq_analytical_amplitude}) and~(\ref{eq_analytical_variance}) into (\ref{eq_asym_gauss_tot}),
the   surface isovector energy can be expressed as a function of 
the symmetry energy coefficients $(J_{sym}, L_{sym}, K_{sym})$:
\begin{widetext}
\bea
E_s^{IV} & = &
3
\sqrt{   \frac{ \pi }  {1-\frac{K_{1/2}}{18 J_{1/2}} }   } 
\frac{\rho_{sat}(0)}{\rho_{sat}(\delta)}
\frac{a_0}{r_{sat}(\delta)} J_{1/2}
\nonumber \\
&\times & \left\{
\frac{1}{4}  \left( \frac{\Delta R(a)}{a(A,\delta)} \right)^2
+ \left[
						\frac{\Delta R(a)}{a(A,\delta)} - \frac{1}{2} \left( \frac{\Delta R(a)}{a(A,\delta)} \right)^2
		 			\right] \delta
 +   					 \left[
						 \left( 1-\frac{J_{sym}}{J_{1/2}} \right)  - \frac{\Delta R(a)}{a(A,\delta)} 
			- \frac{1}{4} \left( 1+ \frac{L_{sym} L_{1/2}}{J_{1/2}K_{sat}} \right) \left( \frac{\Delta R(a)}{a(A,\delta)} \right)^2
						\right] \delta^2
\right\}
\nonumber \\
		& \times &
		\left\{
		A^{2/3} 
		+ \frac{ 2 }{1-\frac{K_{1/2}}{18 J_{1/2}} } \left(  \frac{ a_0}{r_{sat}(\delta) }  \right)^2
		- \frac{2\pi^2}{3} \left( \frac{a(A,\delta)}{r_{sat}(\delta)} \right)^2
		\right\}
.
\label{eq_IV_gauss_final}
\eea
\end{widetext}

In principle  the surface coefficients $(J_{1/2}, L_{1/2}, K_{1/2})$
can be expressed as a function of the bulk ones $(J_{sym}, L_{sym}, K_{sym})$ by using polynomial expansion in the density. 
However,  we can see from Eq.~(\ref{eq_IV_gauss_final})  that the surface isovector energy $E_s^{IV}$
is proportional to the symmetry energy $J_{1/2}$ evaluated at the surface $R$.
It is quite natural that the surface energy component is mainly determined by the surface properties of the nuclei,
and therefore, the surface symmetry energy is mainly proportional to the isovector parameter $J_{1/2}$.
For this reason, expressing Eq.~(\ref{eq_IV_gauss_final}) only in terms of bulk quantities $(J_{sym}, L_{sym}, K_{sym})$
would   make Eq.~(\ref{eq_IV_gauss_final})  less transparent.

For completely symmetric nuclei, that is $\Delta R=0$ and $\delta=0$, the isovector energy is  identically zero as it should.
However, if we neglect the neutron skin thickness only, that is we consider $\Delta R=0$ but $\delta \neq 0$, a non-zero isovector surface energy
is obtained, given by
\bea
E_{surf}^{IV, \Delta R=0} & = &
3
\sqrt{   \frac{ \pi }  {1-\frac{K_{1/2}}{18 J_{1/2}} }   } 
\frac{\rho_{sat}(0)}{\rho_{sat}(\delta)}
\frac{a_0}{r_{sat}(\delta)} 
						 \left( J_{1/2}-J_{sym} \right)
						 \delta^2	A^{2/3} 
.
\nonumber \\
\label{eq_IV_gauss_skin0}
\eea
This expression is proportional to the energy density  difference between bulk and surface $\left(J_{1/2}-J_{sym}\right)$, that is to the $L_{sym}$ parameter.
In this approximation, the diffuseness $a(A,\delta)$ does not appear, which means
that the isovector surface energy  contributes to the determination of the diffuseness  only if we consider the neutron skin.

From a mathematical point of view we can also consider the limit $\delta=0$, $\Delta R \neq 0$, giving:
\bea
E_{surf}^{IV, \delta=0} & = &
\frac{3}{4} 
\sqrt{   \frac{ \pi }  {1-\frac{K_{1/2}}{18 J_{1/2}} }   }  J_{1/2}
  \frac{\Delta R^2(a_0)}{a_0r_{sat}(0)}  A^{2/3}
. 
\label{eq_IV_gauss_delta0}
\eea
This expression shows that an isovector surface energy can be induced in asymmetric nuclei even if no asymmetry is present in the bulk.
Of course in realistic situations the bulk asymmetry and the difference between neutron and proton radii are not independent variables; 
in particular the skin is negligeable if $\delta=0$ as we have already assumed in order to obtain Eq.~(\ref{a0}) above.

Eq.~(\ref{eq_IV_gauss_final}) shows than even in our rather crude approximation the surface symmetry energy presents a very complex 
dependence on the physical quantities that measure  isospin inhomogeneity, namely 
the bulk asymmetry $\delta$ and the neutron skin thickness $\Delta R$.
In particular we find that $E_s^{IV}(A,\delta)$ is not quadratic with $\delta$ but has non-negligible linear components 
(see also Fig.~\ref{Fig_Gauss_IS_IV} below). We have also quantitatively tested that  both  linear and quadratic terms in $\Delta R$
are required to correctly reproduce the   surface isovector energy.
It is interesting to notice that the linear components mix  $\delta$ and $\Delta R$.
Indeed, as we can see in Eqs.~(\ref{eq_IV_gauss_skin0}) and (\ref{eq_IV_gauss_delta0}),
putting to zero one of those variables, which both measure the isospin inhomogeneities,  leads to a quadratic behavior with respect to 
the other variable (cf. eqs~(\ref{eq_IV_gauss_skin0}) and~(\ref{eq_IV_gauss_delta0})).
%
\begin{figure}[htbp]
\includegraphics[angle=270,width=\linewidth, clip]{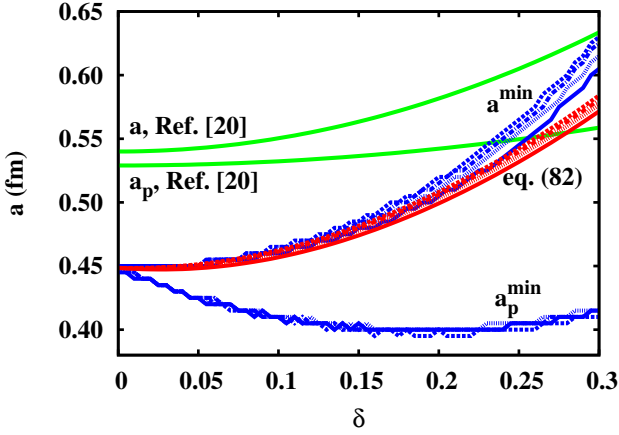}
\caption{
(Color online)
Diffuseness  as a function of the bulk isospin asymmetry.
Red lines: Eq.~(\ref{eq_diffuseness_gauss}) from the minimization of the   gaussian approximation . 
Blue lines: minimization of the exact numerically calculated ETF surface energy.   
Green lines:  fit from HF density profiles, taken from~\cite{pana}.
}
\label{Fig_asym_diffuseness}
\end{figure}

Similar to the previous section, the diffuseness is the only unconstrained parameter of the model. It can therefore
 be determined in a variational approach by minimizing the total (isoscalar and isovector) surface energy.
In section~\ref{sec_sym_diffuseness}, we have shown that only the dominant $\propto A^{2/3}$ terms are important to evaluate the diffuseness.
For this reason, we neglect again terms beyond plane surface,
and we approximate the neutron skin thickness $\Delta R$ by  the hard sphere approximation $\Delta R_{HS}$.
Neglecting the quadratic terms in the expansion in $\Delta R_{HS}/a$, we obtain
\bea
a^2(A,\delta)
& = &
		 a^2_{IS}(\delta)
\nonumber \\ & + &
		 \sqrt{   \frac{ \pi }{1-\frac{K_{1/2}}{18 J_{1/2}} }   }  
		 \frac{\rho_{sat}(0)}{\rho_{sat}(\delta)} 
		 \frac{ 3 J_{1/2} \left( \delta - \delta^2 \right)}{ \mathcal{C}_{surf}^{L}(\delta) }
		 a_0 \Delta R_{HS}(A,\delta)
,
\nonumber \\
\label{eq_diffuseness_gauss}
\eea
where $a_{IS}(\delta)$ is the diffuseness obtained in section~\ref{sec_asym_density_profile} by neglecting 
the isovector component :
$a_{IS}(\delta) = \sqrt{ \mathcal{C}_{surf}^{NL}(\delta) / \mathcal{C}_{surf}^{L}(\delta) }$.
We  found in section~\ref{sec_asym_density_profile} that $a_{IS}$ 
slightly decreases with the isospin asymmetry (see Fig~\ref{Fig_sym_diffuseness_isoscal_deltar}),
which does not appear consistent with the behavior observed in full HF calculations.
Now considering in the variational principle the isovector term in addition to the isoscalar one,
the diffuseness $a$ given by Eq.~(\ref{eq_diffuseness_gauss}) acquires an additional term  
which modifies its global $\delta$ dependence.
The complete result Eq.~(\ref{eq_diffuseness_gauss}) is displayed in Figure~\ref{Fig_asym_diffuseness}.
We can see that the additional term due to the isovector energy contribution 
inverses the trend found section~\ref{sec_asym_density_profile}, as expected. 
More specifically, though it does not clearly appear in Eq.~(\ref{eq_diffuseness_gauss}), 
the analytical diffuseness is seen to quadratically increase with $\delta$,
corroborating the assumption found in Ref.~\cite{pana}.

Although we only considered  terms $\propto A^{2/3}$, as in a slab geometry, 
the results slightly depend on the nucleus mass as shown by the slight dispersion of the different red curves 
in Figure~\ref{Fig_asym_diffuseness}. 
This is due to the neutron skin since $\Delta R_{HS}(A,\delta)$ increases with decreasing mass number $A$.
For comparison, the diffusenesses $a$ and $a_p \neq a$ obtained by a fit of HF density profiles 
in Ref.~\cite{pana} are also represented in Figure~\ref{Fig_asym_diffuseness} (green curves),
as well as the numerically calculated pair $(a^{min},a_p^{min})$ which minimises the energy (blue curves).

\begin{figure}[htbp]
\includegraphics[angle=270,width=\linewidth, clip]{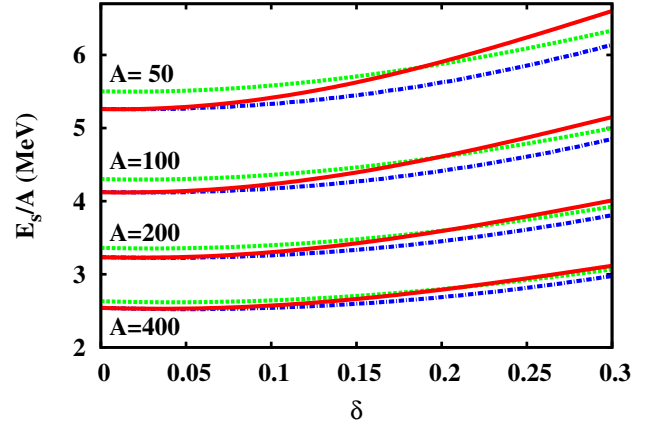}
\caption{
(Color online)
Lower panel: surface energy per nucleon as a function of the bulk isospin asymmetry for four isobaric nucleus chains.  
Full red lines:  gaussian approximation using the diffuseness Eq.~(\ref{eq_diffuseness_gauss}). 
Dash-dotted blue lines: exact numerically calculated ETF surface energy using the optimal diffusenesses $(a^{min},a_p^{min})$ (see text).
Dashed green lines: exact numerically calculated ETF surface energy using the diffusenesses from~\cite{pana}.
}
\label{Fig_asym_diffuseness_E}
\end{figure}

As we can see, these  diffusenesses significantly differ from each other,
but their consequence on the energy is small as we can observe in Fig.~\ref{Fig_asym_diffuseness_E}
which displays the  corresponding   surface energy $E_s=E_s^{IS}+E_s^{IV}$ per nucleon, 
for different isobaric chains.  
In this figure, the blue  curves correspond to a numerical integration of the ETF energy density, using
the diffusenesses  which minimize the total surface energy. These results can thus be considered 
as "exact" ETF results. The use of the very different $a$ and $a_p$ values fitted from HF (green lines) leads to only slightly
different energies, except for the lightest isobar chain.  
The analytical approximation given by the sum of Eq.~(\ref{eq_asym_Enbsym}) and Eq.~(\ref{eq_IV_gauss_final}),
is also plotted (red curves), where the diffuseness is given by the analytical formula Eq.~(\ref{eq_diffuseness_gauss}).
We can see that our analytical approximation closely follows the "exact" ETF results.
 
All the curves show a positive surface symmetry energy,
which contrasts with Fig.~\ref{Fig_apprx_TK_E}.
As it has been discussed in~\cite{aymard}, this change of sign is due to the choice
between the bulk asymmetry $\delta$ or the global asymmetry $I$, in the definition of the bulk energy. 
This choice obviously affects the residual part of the energy $E_s$, since the sum of the two gives the same ETF functional.
This residual part is, to first order, given by the surface symmetry energy as discussed in Ref.~\cite{aymard}.

In order to further validate the analytical results of this section, 
quantitative comparisons with Hartree-Fock calculations are shown in the next section~\ref{sec_IV_compHF}.
%
%
%
%

%
%
%
%
\subsubsection{Comparison to Hartree-Fock calculations}
\label{sec_IV_compHF}
In this section, we explore the level of accuracy of both the no-skin approximation and the gaussian approximation, 
respectively developed in sections~\ref{sec_asym_TK} and \ref{sec_asym_gauss_tot}.

As previously discussed, the two different approximations lead to two different bulk energetics.
Neglecting isospin inhomogeneities implies that the bulk asymmetry $\delta$ is equalized to the average asymmetry $I$.
Thus the bulk quantities $\rho_{sat}$ and $E_b$ defined by Eqs.~(\ref{eq_asym_rho0}) and (\ref{eq_asym_bulk})
depend on $I$, and  the total energy of a nucleus $(A,I)$ within the no-skin approximation is given by
\bea
E_{NoSkin}(A, I) = E_b(A,I) + E_s^{IS}(A,I) + E_s^{IV}(A,I),
\label{eq_Etot_no_skin}
\eea
where $E_s^{IS}(A,I)$ is given by Eq.~(\ref{eq_asym_Enbsym}) (with $I$ instead of $\delta$),
$E_s^{IV}(A,I)$ by Eq.~(\ref{eq_noSkin_IV}),
and the diffuseness  is given by Eq.~(\ref{eq_asym_diffuseness_SLAB_noSkin}).

On the other hand, the gaussian approximation allows defining two independent density profiles.  
Therefore, the bulk energy depends on the bulk asymmetry $\delta(A,I)$ defined by Eq.~(\ref{eq_asym_deltar})
and the total energy of a nucleus $(A,I)$ within this approximation is given by
\bea
E_{Gauss}(A,I) = E_b(A,\delta) + E_s^{IS}(A,\delta) + E_s^{IV}(A,\delta),
\label{eq_Etot_Gauss}
\eea
where $E_s^{IS}(A,\delta)$ is given by Eq.~(\ref{eq_asym_Enbsym}),
$E_s^{IV}(A,\delta)$ by Eq.~(\ref{eq_IV_gauss_final}),
and the isoscalar diffuseness is given 
by Eq.~(\ref{eq_diffuseness_gauss}).

\begin{figure}[htbp]
\includegraphics[angle=270,width=\linewidth, clip]{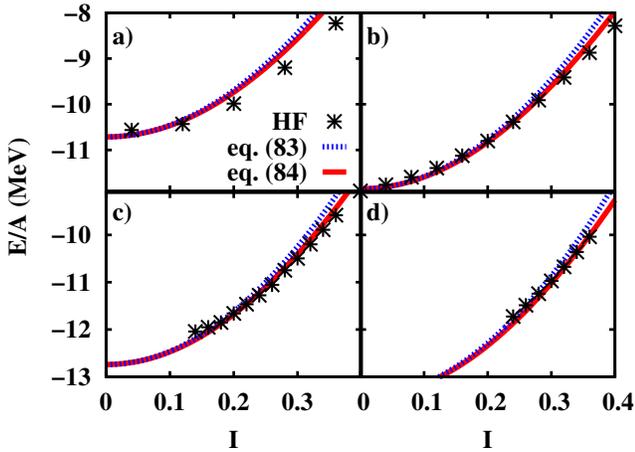}
\caption{
(Color online)
Total energy $E=E_b+E_s$ per nucleon as a function of the nucleus asymmetry $I=1-2Z/A$
calculated within the no-skin approximation, Eq.~(\ref{eq_Etot_no_skin}) (blue dotted lines) 
and within the gaussian approximation, Eq.~(\ref{eq_Etot_Gauss}) (red full lines),
compared to nuclear Hartree-Fock energy (stars).
a) $A=50$; b) $A=100$; c) $A=200$; d) $A=400$.
}
\label{Fig_tot_energy_comparison_HF}
\end{figure}

In Figure~\ref{Fig_tot_energy_comparison_HF},
we compare the analytical expressions~(\ref{eq_Etot_no_skin}) and (\ref{eq_Etot_Gauss}) 
with Hartree-Fock energy calculations, for different isobaric chains.
To compare the same quantities, we used the same interaction (SLy4),
and we have removed the Coulomb energy from the total HF energetics.

We can see from the figure that the no-skin and the gaussian approximations predict close values for the total energy.
For low asymmetries $I \lesssim 0.2$ where the two models are almost undistinguishable:
they reproduce the microscopic calculations with a very good accuracy,
especially for medium-heavy nuclei $A \gtrsim 100$.
However, for higher asymmetries $I \gtrsim 0.2$ where the symmetry energy  becomes important,
a systematic difference between the two models appears 
and increases up to $\sim 400~\mathrm{keV}/A$ for the highest asymmetries $I \sim 0.4$:
the gaussian approximation is systematically  closer to the microscopic results than the no-skin model.
This observation highlights the importance of taking into account the isospin asymmetry inhomogeneities,
 considering the neutron skin and at the same time
differentiating the bulk asymmetry $\delta$ from the global one $I$, as it has been discussed in Ref.~\cite{aymard}.
Quantitatively, for medium-heavy nuclei, the accuracy of Eq.~(\ref{eq_Etot_Gauss}) is better than $\sim 200~\mathrm{keV}/A$,
which is similar to the predictive power of spherical Hartree-Fock calculations for this effective interaction, with respect to experimental data.

To conclude, the gaussian approximation developed in section~\ref{sec_asym_gauss_tot}
provides a reliable analytical formula, especially for the   surface symmetry energy. 
For this reason we will only use the gaussian approximation to further 
study the different components of the nuclei energetics, as we turn to do in the next section.
\section{Study of the different energy terms}
\label{seq_applications}

In this section, we use the analytical formulae
based on the gaussian approximation detailed in section~\ref{sec_asym_gauss_tot},
to study the different components of nuclear energetics.
As we have previously discussed throughout this paper,
we can decompose the nucleus total energy $E$ into bulk $E_b$ and   surface $E_s$ parts.
Both can be written as sums of isoscalar $E_i^{IS}$, that is the part independent of $\rho_3(r)$, and isovector $E_i^{IV}$ terms.
The  surface energy can be further split into plane surface $E_{surf}\propto A^{2/3}$, curvature $E_{curv}\propto A^{1/3}$ and mass independent $E_{ind}$ terms.
Finally, we can distinguish the local $E_i^{IS,L}$ and the non-local $E_i^{IS,NL}$ components of the   surface isoscalar part only, 
since we did not discriminate them in the gaussian approximation used for the isovector energy.
In summary, the energy of a $(A,I)$ nucleus can be written as
\bea
E(A,I) & = & E_b(A,\delta) + E_s(A,\delta), 
\label{eq_decomposition_bulk_s}
\\
E_s(A,\delta) & = & E_s^{IS}(A,\delta) + E_s^{IV}(A,\delta),
\label{eq_decomposition_IS_IV}
\\
E_s^{IV}(A,\delta) & = & E_{surf}^{IV}(A,\delta) + E_{ind}^{IV}(A,\delta),
\label{eq_decomposition_IV_surf_curv}
\\
E_s^{IS}(A,\delta) & = & E_{surf}^{IS}(A,\delta) + E_{curv}^{IS}(A,\delta) + E_{ind}^{IS}(A,\delta),
\label{eq_decomposition_IS_surf_curv}
\\
E_{surf}^{IS}(A,\delta) & = & E_{surf}^{IS,L}(A,\delta) + E_{surf}^{IS,NL}(A,\delta),
\label{eq_decomposition_IS_surf_L_NL}
\\
E_{curv}^{IS}(A,\delta) & = & E_{curv}^{IS,L}(A,\delta) + E_{curv}^{IS,NL}(A,\delta),
\label{eq_decomposition_IS_curv_L_NL}
\eea
where the bijective relation (for a given mass) between $I$ and $\delta$ is given by Eq.~(\ref{eq_asym_deltar}).
The different isoscalar terms $E_i^{IS,j}$ are defined by Eqs.~(\ref{eq_asym_Enbsym}) to (\ref{eq_asym_3D_Es_cl_ind}),
with the diffuseness $a(A,\delta)$  determined within the gaussian approximation, Eq.~(\ref{eq_diffuseness_gauss}).
The isovector components $E_i^{IV}$ are introduced in Eq.~(\ref{eq_IV_gauss_final}),
where the curvature term, in this gaussian approximation, is identically zero by construction.

In the following, we will study each of these terms, and specifically their dependence with the asymmetry $\delta$.
For this comparison, we have chosen a representative isobaric chain  $A=100$ for which the ETF approximation was successfully compared to HF results 
in Fig.~\ref{Fig_tot_energy_comparison_HF}, for the SLy4 interaction.
For this choice of mass, $\delta \approx 0$ corresponds to the proton dripline and $\delta \approx 0.3$ the neutron dripline (see Fig.~\ref{Fig:limits_delta}).

Due to our limited experimental knowledge of the isovector properties of the effective interaction,
the behavior of the different energy terms with asymmetry is to some extent model dependent.
In order to sort out general trends we have considered different Skyrme functionals which approximately span
the current uncertainties on the density dependence of the symmetry energy.

The corresponding bulk parameters are reported in Table~\ref{table:NNparam}.
In this table, the calculated surface coefficients $(J_{1/2},L_{1/2},K_{1/2})$ 
entering Eq.~(\ref{eq_IV_gauss_final}) and (\ref{eq_diffuseness_gauss}) are also given.
As it is well known~\cite{esym_book}, the different interactions are very close at half saturation density, reflecting the fact that 
all Skyrme parameters have been fitted on ground state properties of finite nuclei, which correspond to an average density of the order
of $\rho_{sat}/2$. Nevertheless, a considerable spread is already seen at saturation density, showing that the extrapolation of isovector
properties to unexplored density domains is still not well controlled~\cite{esym_book}.

Concerning the LNS interaction, the parametrization proposed in Ref.~\cite{lns} corresponds to a too high saturation density 
which is not realistic. This induces a trivial  deviation with respect to the other interactions in both the bulk and surface isovector components.
For this reason, only the isovector properties  of this functional are of interest for this study.

A more complete study of the effective interactions parameter space would be necessary
to reach sound conclusions on the quantitative model dependence, 
but from the representative chosen interactions, we can already dress some
qualitative interpretations.

\begin{table*}[tbph]
\setlength{\tabcolsep}{.14in}
\renewcommand{\arraystretch}{1.1}
\caption{
Bulk and surface nuclear properties for the different Skyrme interactions examined in this paper.
}
\begin{tabular}{c|cccccccccc}
\hline
$~$                   && $\rho_{sat}(0)$ &$m^*/m$ & $K_{sat}$ & $J_{sym}$ & $L_{sym}$    & $K_{sym}$       & $J_{1/2}$  & $L_{1/2}$  & $K_{1/2}$ \\
Interaction           && (fm$^{-3}$)   &  & (MeV)     & (MeV)     & (MeV)        & (MeV)           & (MeV)      & (MeV)      & (MeV) \\
\hline
SLY4~\cite{sly4}      && 0.1595     & 0.595     & 230.0     &  32.00    & \hfill  46.0 & \hfill  -119.8  & 22.13      &  38.6      &  -74.0 \\
SkI3~\cite{ski3}      && 0.1577      &0.577     & 258.2     &  34.83    & \hfill 100.5 & \hfill    73.0  & 18.85      &  46.7      &  -25.2 \\
SGI~\cite{sgi}        && 0.1544       &0.608     & 261.8     &  28.33    & \hfill  63.9 & \hfill  - 52.0  & 16.75      &  38.4      &  -29.7 \\
LNS~\cite{lns}        && 0.1746      &0.826     & 210.8     &  33.43    & \hfill  61.5 & \hfill  -127.4  & 21.10      &  44.6      &  -56.8 \\
\end{tabular}
\label{table:NNparam}
\end{table*}

\begin{figure}[htbp]
\includegraphics[width=0.9\linewidth, clip]{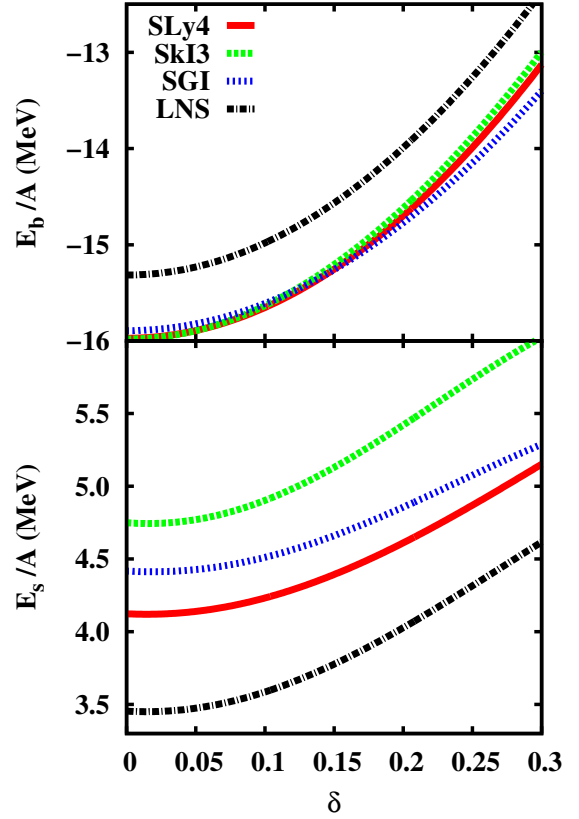}
\caption{
(Color online) 
Bulk (upper panel) and   surface (lower panel) energy per nucleon
as a function of the bulk asymmetry $\delta$ for isobaric nuclei $A=100$,
predicted by Eq.~(\ref{eq_Etot_Gauss}).
Different Skyrme interactions are considered:
SLy4~\cite{sly4} (full red), 
SkI3~\cite{ski3} (dashed green), 
SGI~\cite{sgi} (dotted blue), 
LNS~\cite{lns} (dashed-dotted black).
}
\label{Fig_Gauss_bulk_surf}
\end{figure}

The bulk energy  per nucleon is shown in the upper panel of Fig.~\ref{Fig_Gauss_bulk_surf}.
At low asymmetries, the curves are indistinguishable reflecting the good present knowledge of symmetric nuclear matter properties.
The only exception is given by LNS, which presents a global shift with respect to the other functionals.
As already remarked, this is due to the irrealistically high saturation density of this parametrization (tab.~\ref{table:NNparam}).
However, we can see that the behavior with isospin is comparable to the one of the other functionals, reflecting a compatible bulk symmetry energy.   
For the highest asymmetries $\delta \gtrsim 0.25$, we can see that all the parametrizations differ,
which reflects the larger uncertainties for asymmetric matter.

The lower panel of Figure~\ref{Fig_Gauss_bulk_surf} displays the   surface corrections.
We can see that the qualitative behaviour of the different models is the same: 
 $E_s/A$ increases with the asymmetry, leading to a positive sign of the corresponding symmetry energy.
As it has been already discussed in Ref.~\cite{aymard}, 
this comes from the consideration of the bulk asymmetry $\delta$  instead of the global one $I$ in the definition of the nuclear bulk.

The increase rate with isospin is not the same in the different models, reflecting the different surface symmetry energies of the functionals.
In particular, the   steep behaviour predicted by the SkI3 parametrization
is due to the  stiff isovector properties of this effective interaction 
(see $L_{sym}$ and $K_{sym}$ in tab.~\ref{table:NNparam}), which lay close to the higher border of the presently accepted values for these 
parameters\cite{esym_book}.

Moreover, the four considered interactions predict very different values of $E_s$.
In particular, at $\delta = 0$ for which the SLy4, SkI3 and SGI models are in perfect agreement on the bulk energy,
they however differ from $\sim 500$~keV per nucleon on the   surface energies.
We will come back to this surprising result later in this section.

\begin{figure}[htbp]
\includegraphics[width=0.9\linewidth, clip]{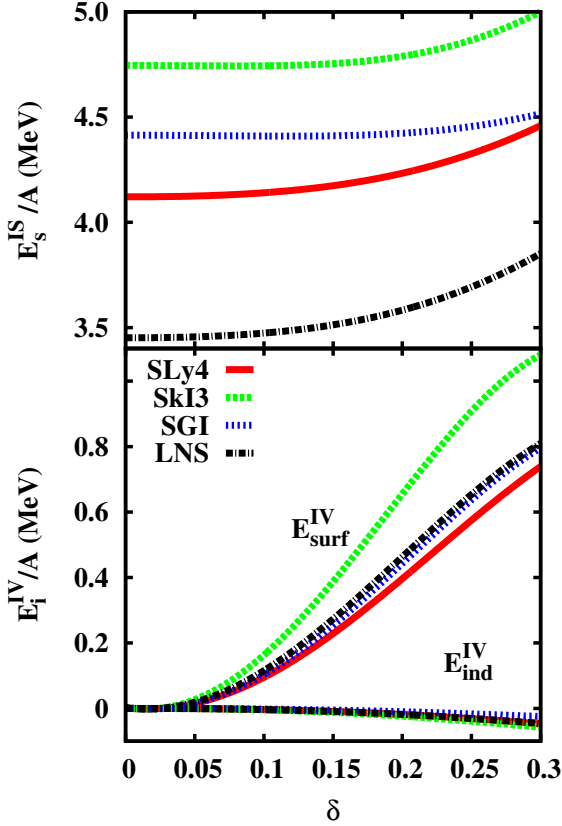}
\caption{
(Color online)
Isoscalar  (upper panel) and isovector (lower panel) surface energy per nucleon 
as a function of the bulk asymmetry $\delta$ for isobaric nuclei $A=100$,
predicted by Eq.~(\ref{eq_Etot_Gauss}).
Different Skyrme interactions are considered:
SLy4~\cite{sly4} (full red), 
SkI3~\cite{ski3} (dashed green), 
SGI~\cite{sgi} (dotted blue), 
LNS~\cite{lns} (dashed-dotted black).
}
\label{Fig_Gauss_IS_IV}
\end{figure}

Fig.~\ref{Fig_Gauss_IS_IV} shows the energy decomposition of
Eqs.~(\ref{eq_decomposition_IS_IV}) and (\ref{eq_decomposition_IV_surf_curv}).
As expected, at $\delta=0$, though not identically zero (see Eq.~(\ref{eq_IV_gauss_delta0})), 
the isovector energy (lower panel) is completely negligible. This a-posteriori justifies the assumption $E_s^{IV}(0)=0$ we made 
in order to obtain $a_0$ in Eq.~(\ref{eq_analytical_variance}).
However, for asymmetric systems, though  smaller than the isoscalar energy (upper panel), the isovector energy cannot be neglected.
Indeed, its dependence with $\delta$ is much stronger, meaning that the isovector term is the most important term determining the surface symmetry energy . Concerning  the mass independent term, we can see that it is negligible  compared to the other components,
 as expected for the medium-heavy nucleus concerned by this picture.
Finally, we can observe that the isovector energy is not quadratic with $\delta$, thus confirming  that the linear terms of Eq.~(\ref{eq_IV_gauss_final}) cannot be neglected.

\begin{figure}[htbp]
\includegraphics[width=0.9\linewidth, clip]{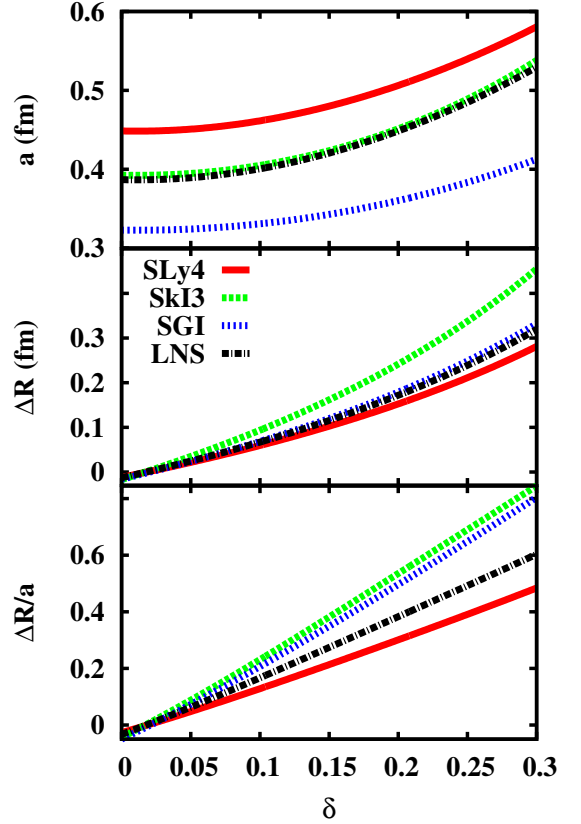}
\caption{
(Color online)
Diffuseness $a$ (upper panel), neutron skin thickness $\Delta R$ (middle panel) and the ratio $\Delta R /a$ (lower panel)  
as a function of the bulk asymmetry $\delta$ for the isobaric chain $A=100$, predicted within the gaussian approximation (see text).
Different Skyrme interactions are considered:
SLy4~\cite{sly4} (full red), 
SkI3~\cite{ski3} (dashed green), 
SGI~\cite{sgi} (dotted blue), 
LNS~\cite{lns} (dashed-dotted black).
}
\label{Fig_Gauss_a_DeltaR}
\end{figure}

Fig.~\ref{Fig_Gauss_a_DeltaR} shows the predictions of the different functionals concerning the parameters associated to the density profiles, namely 
the diffuseness (upper panel), the neutron skin (middle panel) and their ratio (lower panel).
We can see that, for a given asymmetry $\delta$, the spread of the diffuseness  values given by Eq.~(\ref{eq_diffuseness_gauss}) is very important,
reflecting the poor knowledge of this quantity.
These large uncertainties  can be understood considering  that the
diffuseness does not seem to affect the energy in a systematic way.
In particular, though SkI3 and SGI models surprisingly give the same diffuseness, the corresponding surface properties systematically differ.
Moreover, this similarity of the diffuseness cannot be straightforwardly linked 
to any specific interaction property or parameter (see tab.~\ref{table:NNparam}).
This reflects again the fact that the diffuseness is a delicate balance of all energy components,
and is determined by very subtle competing and opposite effects.

The middle part of the Figure shows the obvious correlation between $\Delta R=R-R_p$ and $\delta$. 
It is clear from this behavior that quadratic terms in the neutron thickness cannot be neglected to correctly estimate
the symmetry energy (see Eq.~(\ref{eq_IV_gauss_final})).
It is interesting to observe that the SGI and LNS models give very close results for this quantity, and the same was true 
for the isovector part of the surface energy in Figure~\ref{Fig_Gauss_IS_IV} above.

This comes from the  fact, already observed in the literature \cite{esym_book}, that $\Delta R$ is mainly determined by the slope of the symmetry energy $L_{sym}$~\cite{esym_book}
which are close in the SGI and LNS models.
Our work confirms that the neutron thickness can be viewed as a measurement of the $L$ parameter.
Indeed,   $\Delta R$ can be well approximated using  the equivalent hard spheres radii $R_{HS}(\delta)$, $R_{HS,p}(\delta)$, see Eq.~(\ref{eq_neutron_skin_thickness}).
This means that  $\Delta R$ can be seen as a function of  the saturation density $\rho_{sat}(\delta)$. In turn, the saturation density is given by Eq.~(\ref{eq_asym_rho0}) which at first order is quadratic in $\delta^2$ with the coefficient $L_{sym}/K_{sat}$. 
Since $K_{sat}$ is relatively well constrained, we then understand why $\Delta R$ is  mainly determined by $L_{sym}$.
 In particular, the neutron skin thickness is predicted to be the same in the two specific interactions SGI and LNS.
Since the surface isovector energy Eq.~(\ref{eq_IV_gauss_final}) at a given bulk asymmetry mainly depends on the neutron skin, 
this also explains why we obtain the same energies for the two models in Fig.~\ref{Fig_Gauss_IS_IV}.

This essential role of $\Delta R$  to determine the symmetry energy is confirmed observing 
from Fig.~\ref{Fig_Gauss_IS_IV} and \ref{Fig_Gauss_a_DeltaR}
that Skyrme models which predict thicker neutron skin, that is higher $L_{sym}$, give systematically larger values of the isovector surface energy.

The lower part of Figure~\ref{Fig_Gauss_a_DeltaR} shows 
the ratio $\Delta R /a$ as a function of $\delta$.
Though it is the quantity which mainly governs the behavior of Eq.~(\ref{eq_IV_gauss_final}),
it does not constrain the surface isovector energy $E^{IV}$.
Indeed, same $\Delta R /a$ from the functionals SkI3 and SGI lead to different energies (Fig.~\ref{Fig_Gauss_IS_IV}, lower panel),
corroborating the above discussion: 
only the $L$ parameter, or equivalently the neutron skin thickness $\Delta R$, is relevant to determine the isovector contribution.

This stresses the importance of the experimental  measurement of neutron skin thickness as a key quantity for the knowledge of the 
density dependence of the symmetry energy \cite{esym_book}.

\begin{figure}[htbp]
\includegraphics[angle=270,width=\linewidth, clip]{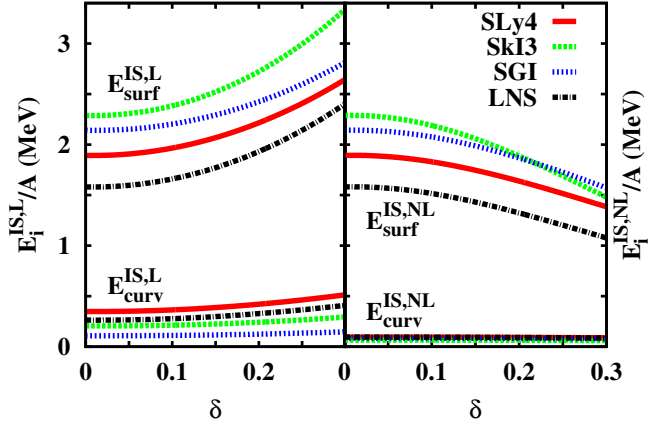}
\caption{
(Color online)
Decomposition of the
local (left) and non-local (right) part of the isoscalar   surface energy per nucleon, into its surface and curvature 
component 
as a function of the bulk asymmetry $\delta$ for the isobaric chain $A=100$,
as predicted by Eq.~(\ref{eq_Etot_Gauss}).
Different Skyrme interactions are considered:
SLy4~\cite{sly4} (full red), 
SkI3~\cite{ski3} (dashed green), 
SGI~\cite{sgi} (dotted blue), 
LNS~\cite{lns} (dashed-dotted black).
}
\label{Fig_Gauss_L_NL}
\end{figure}

To conclude, we study in Figure~\ref{Fig_Gauss_L_NL} the decomposition into  local and non-local terms as predicted by the different functionals.
Only the isoscalar part of the surface energy is considered because 
these different terms are mixed up in the gaussian approximation we have employed for the isovector component.

Again, we can see that the qualitative behavior of the different Skyrme models is the same for each specific term.
We can then safely conclude that the non-local curvature component $E_{curv}^{IS,NL}$
can be neglected for medium-heavy nuclei $A \gtrsim 100$, but the local curvature energy has to be taken into account
since it represents for these nuclei $10$\% to $25$\% of the total surface local energy,
depending on the interaction choice and on the asymmetry $\delta$.

Concerning the $\delta$ dependence of the isoscalar surface energies in Fig.~\ref{Fig_Gauss_L_NL},
we can notice that the local and non-local parts have opposite behaviors, 
leading to the rather flat curves observed in Fig.~\ref{Fig_Gauss_IS_IV}, upper panel.
In section~\ref{sec_sym_results} and \ref{sec_asym_non_bulk_energy}, we have  shown that 
 the exact equality $E_{surf}^{IS,L}=E_{surf}^{IS,NL}$ (Eq.(\ref{magic})) is obtained  only
if both curvature and isovector terms are neglected in the determination of the diffuseness.
 However, the neglect of isovector terms leads to  a wrong dependence with $\delta$ as shown in Fig.~\ref{Fig_sym_diffuseness_isoscal_deltar}.
 Thus, isovector terms cannot be avoided.

The results of Figure~\ref{Fig_Gauss_L_NL} clearly show that, once these terms are consistently added 
in the variational procedure (Eq.~(\ref{eq_diffuseness_gauss})),   the equality $E_{surf}^{IS,L}=E_{surf}^{IS,NL}$
 is completely violated for asymmetric systems.
Therefore the isoscalar energy strongly depends on the neutron skin thickness, even if it is an indirect 
dependence through the diffuseness.
This shows that, though the energy can be splitted into different terms, these latter cannot be decorrelated
and have to be treated altogether.

We have already observed in Figure~\ref{Fig_Gauss_bulk_surf} that the different functionals 
predict very different surface energy at $\delta=0$, which might be surprising considering that the symmetric nuclear 
properties are supposed to be well constrained by experimental data.  An obvious interpretation would be that the 
discrepancy comes from the surface properties, that is the non-local gradient terms and the (poorely constrained) 
diffuseness parameter. 
However, comparing  the different values of the predicted diffuseness at $\delta=0$ from Figure~\ref{Fig_Gauss_a_DeltaR},
we can see that 
$a_{SGI}<a_{LNS}=a_{SkI3}<a_{SLy4}$. This inequality sequence is not respected for the surface energy 
$E_{surf}(\delta=0)$ in Fig.~\ref{Fig_Gauss_L_NL}, meaning that the difference of surface energies cannot be ascribed to the diffuseness.

The possible dependence on the couplings of gradient and spin-orbit terms is also excluded. Indeed, 
we can see from Figure~\ref{Fig_Gauss_L_NL} that
at $\delta=0$, the isovector part is zero by definition and therefore the equality $E_{surf}^{IS,L}=E_{surf}^{IS,NL}$ is verified.
This means that the total surface energy for symmetric bulk is $E_{surf} = 2E_{surf}^{IS,L}$, which does not depend on the non-local terms of the functional, but
only depends
on the bulk interaction coefficients $(\rho_{sat}(0),C_0,C_3,C_{eff},\alpha)$ according to Eqs.~(\ref{eq_sym_coeff_L_surf})-(\ref{eq_sym_coeff_L}).

We can conclude that the differences of the total   surface energies  observed 
for $\delta=0$, that is nuclei very close to isospin symmetry, in Figure~\ref{Fig_Gauss_bulk_surf},
does not come from the non-local properties but
are intrinsically linked to the bulk interaction coefficients $(C_0,C_3,C_{eff},\alpha)$, though 
the SLy4, SkI3 and SGI models correpond to compatible isoscalar equations of state (that is: compatible values for the 
 saturation density $\rho_{sat}(0)$, bulk  energy $E_b(\delta=0)$, compressibility $K_{sat}$ and effective mass). 
This shows that, at variance with the skin thickness $\Delta R$ which is strongly correlated to the isovector equation of state, 
 the nuclear surface energy very poorely constrains the equation of state, even for symmetric or quasi-symmetric nuclei.

\section{Summary and conclusions}
\label{sec_conclu}

In this paper we have addressed the problem of the determination of an analytical mass formula with coefficients directly linked to the different parameters of standard Skyrme functionals, in the extended Thomas-Fermi (ETF) approximation at second order in $\hbar$.
The purpose of this effort is twofold. On one side, such a formula is useful for astrophysical applications where extendend calculations are needed
covering the whole mass table and using a variety of effective interaction to assess the sensitivity of astrophysical observables to the nuclear physics inputs~\cite{compose}. On the other side, analytical expressions of the different coefficients of the mass formula in terms of the Skyrme couplings
allow a better understanding of the correlation between these couplings and the different aspects of nuclear energetics, 
for the construction of optimized fitting procedures of the functionals. 

The modelling of Fermi density profiles proposed in Ref.~\cite{pana} allows an (almost) exact analytical evaluation of the isoscalar part of 
the nuclear energy, naturally leading to the appearance in the surface energy of a curvature term and a constant term independent of the baryonic number. The diffuseness of the density profile is variationally calculated within the same formalism, and a simple analytical expression is given. 
The relative importance of local and non-local terms is studied in detail. Non-local energy components arise both from  gradient and spin-orbit  
in the Skyrme functional, and from the higher $\hbar$ terms in the Wigner-Kirkwood expansion of the kinetic energy.
We show that in the limit of semi-infinite matter the isoscalar surface energy is $\propto A^{2/3}$ and solely depends on the local terms.
This remarkable property already observed in Ref.~\cite{krivine} is however violated in finite nuclei even if spherical symmetry is assumed, and both
components contribute in a complex way to the determination of the surface energy. 
However, the huge dispersion observed on the value of the surface tension for symmetric nuclei in modern Skyrme functionals is essentially due to the 
local couplings, even if these different functionals correspond to comparable saturation properties of symmetric nuclear matter.
This finding means that nuclear matter properties are not sufficient to pin down surface properties of finite nuclei even in the symmetric case.

The extension to isospin asymmetric nuclei is highly non-trivial. No exact analytical integration of the ETF functional is possible in the presence
of isospin inhomogeneities, and approximations have to be done.
We have proposed two different approximations for the determination of the surface symmetry energy.
The first approximation consists in completely negecting the difference between the neutron and proton radius, that is the neutron skin $\Delta R$.
The resulting surface energy shows a quadratic dependence on the isospin asymmetry $I$, and consists of local and non-local plane surface, curvature
and mass dependent terms which are simple generalizations of the expressions obtained for symmetric nuclei.
Surprisingly, this crude approximation  reproduces very well numerical Hartree-Fock (HF) results for all stable nuclei up to asymmetries of the order of $I\approx 0.2$, and leads to a relatively limited overestimation of the order of $\approx 400$ KeV/nucleon close to the driplines.

A better approximation is obtained if isospin inhomogeneities are accounted for. To this aim,  we have introduced a different radius for the neutron and proton distributions, as well as an explicit difference between the global asymmetry $I$ and the asymmetry in the nuclear bulk $\delta$, due to both Coulomb 
and neutron skin effects. 
In this more general case, to obtain a mass formula we make the assumption that the surface energy density is peaked at the nuclear surface, and curvature terms can be neglected. A reproduction of HF results within $\approx 200$ KeV/nucleon at the driplines is obtained, and simple expressions are given for the surface energy and the surface diffuseness parameter.
In particular we show that both linear and quadratic terms in $\delta$ and $\Delta R$ are needed to correctly explain the surface term.
Moreover, within this analytical mass formula, we show that the neutron skin is essentially determined by the slope of the symmetry energy at saturation,
thus confirming earlier numerical results from different groups~\cite{esym_book}.
Conversely, the surface symmetry energy is shown to be due to a complex interplay of all different local and non-local terms in the energy functional.
This implies that constraints on the symmetry energy parameters ($J_{sym}$,$L_{sym}$,$K_{sym}$) from mass measurements might be model dependent and misleading.
As a further developement of this work, we plan to extend the mass formula to the case of neutron-rich nuclei beyond the dripline in equilibrium with a neutron (and possibly proton) gas. Such a parametrization will allow including modifications of the nuclear surface energy due the presence of continuum states  in nuclear statistical equilibrium models, currently used for different astrophysical applications in supernova and neutron star physics~\cite{compose}. A self-consistent inclusion of pairing effects in the local density BCS approximation, using consistent calculations for the mean field and gap equation with the same energy functional, is also in progress~\cite{burrello}.

\begin{acknowledgments}
This work has been partially funded by the SN2NS project ANR-10-BLAN-0503 and it has
been supported by New-Compstar, COST Action MP1304.
\end{acknowledgments}
\newpage
%
%
\appendix
\section{The Skyrme effective interaction}
\label{sec_effective_interaction}
The Skyrme functional for the energy density $\mathcal{H}(\mathbf{r})$ is expressed as \cite{Bender2003a,sly4}
\bea
\mathcal{H}(\mathbf{r}) &=& 
\mathcal{K}(\mathbf{r}) +
\mathcal{H}_0(\mathbf{r}) +
\mathcal{H}_3(\mathbf{r}) +
\mathcal{H}_{eff}(\mathbf{r}) + \nonumber \\
&+& \mathcal{H}_{fin}(\mathbf{r}) +
\mathcal{H}_{so}(\mathbf{r}) +
\mathcal{H}_{sg}(\mathbf{r})
,
\label{eq_density_energy_Skyrme}
\eea
where the kinetic term, the effective mass term, the zero-range term, the density-dependent term, the finite-range term, the spin-orbit term and the spin-gradient term are respectively
\begin{align}
\mathcal{K}       & = \frac{\hbar^2}{2m}\tau ,\nonumber\\
\mathcal{H}_{eff} & = C_{eff}\rho\tau + D_{eff}\rho_3\tau_3 ,\nonumber\\
\mathcal{H}_0     & = C_0\rho^2 + D_0\rho_3^2 ,\nonumber\\
\mathcal{H}_3     & = (C_3\rho^2 + D_3\rho_3^2)\rho^\alpha ,\nonumber\\
\mathcal{H}_{fin} & = C_{fin}(\boldsymbol{\nabla}\rho)^2 + D_{fin}(\boldsymbol{\nabla}\rho_3)^2 ,\nonumber\\
\mathcal{H}_{so}  & = C_{so}\mathbf{J}\cdot\boldsymbol{\nabla}\rho + D_{so}\mathbf{J}_3\cdot\boldsymbol{\nabla}\rho_3 ,\nonumber\\
\mathcal{H}_{sg}  & = C_{sg}\mathbf{J}^2 + D_{sg}\mathbf{J}_3^2,
\label{eq_energy_Skyrme_parts}
\end{align}
and where we have introduced the local isoscalar and isovector particle densities, kinetic densities and spin-orbit density vectors:
\begin{align}
\rho(\mathbf{r})         & = \rho_n(\mathbf{r}) + \rho_p(\mathbf{r}) ,\nonumber\\
\rho_3(\mathbf{r})       & = \rho_n(\mathbf{r}) - \rho_p(\mathbf{r}) ,\nonumber\\
\tau(\mathbf{r})         & = \tau_n(\mathbf{r}) + \tau_p(\mathbf{r}) ,\nonumber\\
\tau_3(\mathbf{r})       & = \tau_n(\mathbf{r}) - \tau_p(\mathbf{r}) , \nonumber\\
\mathbf{J}(\mathbf{r})   & = \mathbf{J}_{n}(\mathbf{r}) + \mathbf{J}_{p}(\mathbf{r}) ,\nonumber\\
\mathbf{J}_3(\mathbf{r}) & = \mathbf{J}_{n}(\mathbf{r}) - \mathbf{J}_{p}(\mathbf{r})
.
\label{eq_def_iso_densities}
\end{align}
The coefficients $C_i$  in equations (\ref{eq_energy_Skyrme_parts}),   associated with the isoscalar contribution, are  linear combinations of the traditional Skyrme parameters $t_i$, $x_i$ and $W_0$ as follows:
\begin{align}
C_0      & = \frac{3}{8} t_0, 	\nonumber\\
C_3      & = \frac{1}{16} t_3 ,	\nonumber\\
C_{eff}  & = \frac{1}{16}\big[ 3t_1 + t_2(4x_2+5) \big] ,\nonumber\\
C_{fin}  & = \frac{1}{64}\big[ 9t_1 - t_2(4x_2+5) \big]  ,\nonumber\\
C_{so}   & = \frac{3}{4}W_0, \nonumber\\
C_{sg}   & = \frac{1}{32}\big[ t_1(1-2x_1) - t_2(1+2x_2) \big],
\label{eq_def_CSkyrme_coeff}
\end{align}
while the  $D_i$ coefficients, associated to the isovector part of the energy, are given by:
\begin{align}
 D_0      & = -\frac{1}{8} t_0 \big[ 2x_0+1 \big] ,\nonumber\\
 D_3      & = -\frac{1}{48} t_3 \big[2x_3+1\big], \nonumber\\
 D_{eff}  & = \frac{1}{16}\big[ t_2(2x_2+1) - t_1(2x_1+1) \big] ,\nonumber\\
 D_{fin}  & = -\frac{1}{64}\big[ 3t_1(2x_1+1) - t_2(2x_2+1) \big] ,\nonumber\\
 D_{so}   & = \frac{1}{4}W_0, \nonumber\\
 D_{sg}   & = \frac{1}{32}\big[ t_1 - t_2 \big]
.
\label{eq_def_DSkyrme_coeff}
\end{align}
The semi-classical development in $\hbar$, so-called Extended Thomas-Fermi (ETF),
provides expressions for the kinetic densities and spin-orbit density vectors,
that is at the second order \cite{Brack1985}:
\begin{align}
\tau_q(\mathbf{r})  & =
\tau_{0q}(\mathbf{r}) + \tau_{2q}(\mathbf{r}) + O(\hbar^4)
\label{eq_def_ETF_tau},
\\
\mathbf{J}_{q}(\mathbf{r}) & =
\mathbf{J}_{0{q}}(\mathbf{r}) + \mathbf{J}_{2{q}}(\mathbf{r}) + O(\hbar^4)
.
\label{eq_def_ETF_J}
\end{align}
The results of nuclear matter calculations give the zeroth order and read:
\begin{align}
\tau_{0q}(\mathbf{r}) & =
\frac{3}{5} (3\pi^2)^{2/3} \rho_{q}(\mathbf{r})^{5/3}
\label{eq_ETF_tau0},
\\
\mathbf{J}_{0{q}}(\mathbf{r}) & = 
\mathbf{0}
.
\label{eq_ETF_J0}
\end{align}
The Wigner-Kirkwood expansion gives the second order of the kinetic densities development:
\bea
\tau_{2q}(\mathbf{r}) = \tau_{2q}^l(\mathbf{r}) + \tau_{2q}^{nl}(\mathbf{r}) + \tau_{2q}^{so}(\mathbf{r})
,
\label{eq_def_ETF_2_tau}
\eea
with
\begin{align}
\tau_{2q}^l & =
\frac{1}{36} \frac{\left (\boldsymbol{\nabla}\rho_q \right)^2}{\rho_q}  +  \frac{1}{3}\Delta\rho_q,
\nonumber\\
\tau_{2q}^{nl} & =
\frac{1}{6} \frac{ \boldsymbol{\nabla}\rho_q \boldsymbol{\nabla} f_q}{f_q}
+\frac{1}{6} \rho_q  \frac{\Delta f_q}{f_q}
- \frac{1}{12} \rho_q  \left(  \frac{\boldsymbol{\nabla} f_q}{f_q} \right) ^2,
\nonumber\\
\tau_{2q}^{so} & =
\frac{1}{2} \left( \frac{2m}{\hbar^2} \right) ^2 \rho_q \left( \frac{W_q}{f_q} \right) ^2
.
\label{eq_ETF_2_tau}
\end{align}
The second order of the Thomas-Fermi approximation for the spin-orbit currents $J_{2q}(\mathbf{r})$ reads
\bea
J_{2q} = - \frac{2m}{\hbar ^2} \rho_q \frac{W_q}{f_q} 
,
\label{eq_ETF_2_J}
\eea
where we have introduced the effective mass coefficients $f_q(\mathbf{r})=m/m_q^*(\mathbf{r})$ with $m_q^*(\mathbf{r})$ the effective masses, and the spin-orbit potentials $\mathbf{W}_{q}(\mathbf{r})$ as follows \cite{Bartel2008}:
\begin{align}
f_q                    & =
1 + \frac{2m}{\hbar^2} \left( C_{eff}\rho \pm D_{eff} \rho_3 \right) 
\label{eq_MF_effmass} ,
\\
\mathbf{W}_{q}  & =
C_{so} \boldsymbol{\nabla}\rho \pm D_{so} \boldsymbol{\nabla}\rho_3
\nonumber\\ & \hphantom{=}+
2C_{sg}\mathbf{J} \pm 2D_{sg}\mathbf{J}_3,
\label{eq_MF_pot_SO}
\end{align}
where   $\pm$  stand for  neutrons (protons) .

In several Skyrme interactions (such as SLy4, SIII, SGII...), the spin-gradient term $\mathcal{H}_{sg}$ are neglected.
Therefore in the following, we take $C_{sg}=D_{sg}=0$,
which in particular uncouples the equations (\ref{eq_ETF_2_J}) and (\ref{eq_MF_pot_SO}).
For more general Skyrme interactions, a full treatment of the spin-gradient terms should be implemented \cite{Bartel2008}.

In symmetric matter, we can set the usual following equalities, at every location $\mathbf{r}$
\begin{align}
2\rho_{q}(\mathbf{r}) & = \rho(\mathbf{r}) , \nonumber\\
2\tau_q (\mathbf{r})   & = \tau(\mathbf{r}) = \tau_0(\mathbf{r}) + \tau_2^l(\mathbf{r}) +  \tau_2^{nl}(\mathbf{r})  +  \tau_2^{so}(\mathbf{r}) ,\nonumber\\
2\mathbf{J}_{q}(\mathbf{r}) & =  \mathbf{J}(\mathbf{r})
,
\label{eq_def_sym_rho_tau_J}
\end{align}
with
\begin{align}
\tau_0      & = \frac{3}{5} \left( \frac{3\pi^2}{2} \right)^{2/3} \rho^{5/3}  ,
\nonumber\\
\tau_2^L    & = \frac{1}{36} \frac{ \left( \boldsymbol{\nabla}\rho \right)^2 }{\rho} + \frac{1}{3} \Delta\rho ,
\nonumber\\
\tau_2^{NL} & =
\frac{1}{6} \frac{ \boldsymbol{\nabla}\rho \boldsymbol{\nabla} f}{f}
+\frac{1}{6} \rho  \frac{\Delta f}{f}
- \frac{1}{12} \rho  \left(  \frac{\boldsymbol{\nabla} f}{f} \right) ^2,
\nonumber\\
\tau_2^{so} & =
\frac{1}{2} \left( \frac{2m}{\hbar^2} \right) ^2 \rho \left( \frac{ C_{so}\boldsymbol{\nabla}\rho}{f} \right) ^2,
\label{eq_sym_rho_tau}
\\
J           & = - \frac{2m}{\hbar ^2} \rho \frac{C_{so}\boldsymbol{\nabla}\rho}{f} 
,
\label{eq_sym_rho_J}
\end{align}
where we have used $\mathbf{W}_q=C_{so}\boldsymbol{\nabla}\rho$ and 
where we have introduced the effective mass
\bea
f = f_q = 1 + \kappa\rho \hspace{0.5cm} \text{with} \hspace{0.5cm} \kappa = \frac{2m}{\hbar^2} C_{eff}.
\label{eq_sym_f}
\eea
With these formulae, the energy density given by Eqs.~(\ref{eq_energy_Skyrme_parts}) straightforwardly reads
\bea
\mathcal{H}[\rho] &=&
h(\rho)
+ \frac{\hbar^2}{2m} f (\tau_2^l + \tau_2^{nl})
+ C_{fin} \left( \boldsymbol{\nabla}\rho \right)^2 \nonumber \\ 
&+& V_{so} \frac{\rho}{f} \left( \boldsymbol{\nabla}\rho \right)^2
\label{eq_sym_density_energy_Skyrme_appendix},
\eea
where we have highlighted the local terms
\bea
h(\rho) = 
\frac{\hbar^2}{2m} f \tau_0 + C_0\rho^2 + C_3\rho^{\alpha + 2},
\label{eq_energy_zeroth_loc_sym_appendix} 
\eea
and where we have gathered the spin-orbit current ($C_{so} \mathbf{J} \cdot \boldsymbol{\nabla}\rho$)
and kinetic density ($\frac{\hbar^2}{2m} f \tau_2^{so}$)
terms which lead to 
the definition of the spin-orbit potential $V_{so} = -\frac{1}{2} \frac{2m}{\hbar^2} C_{so}^2$.
%
%
%
%
%
%
%
%
%
%
%
\section{Integrals of Fermi functions}
\label{sec_appendix_int_FD}
We give here the formulae useful to analytically integrate Fermi functions to some power.
\subsection{General formulae}
\label{sec_appendix_sym_formula}
The Fermi function $F(r)$ (Eq.~(\ref{eq_sym_density_profile})) 
to any power can be integrated in any dimension in using the following general formula~\cite{krivine_math}:
\bea
I_{m,\gamma} &=&
4 \pi
\int_0^{+\infty} \mathrm dr F^{\gamma}(r)r^m \\
&\simeq&
4 \pi
\frac{R^{m+1}}{m+1}
	\left[
		1 + (m+1)
				\sum_{k=0}^m \binom{m}{k}  \eta^{(k)}_\gamma \left( \frac{a}{R} \right)^{k+1}
	\right] \nonumber
,
\label{eq_appendix_Imunu}
\eea
with $m \in \mathbb{N}$, $\gamma \in \mathbb{R^{+*}}$,
\bea
\eta^{(k)}_{\gamma} =
(-1)^k
\int_0^\infty \mathrm du
\left[
	\frac{1+ (-1)^k \e^{-\gamma u}}{\left(1+\e^{-u}\right)^\gamma} 
	-1
\right]
u^k,
\label{eq_appendix_gen_eta}
\eea
and the binomial coefficient $\binom{m}{k}=m!/(k!(m-k)!)$.
The values of the coefficients that have been used for this work are given in table~(\ref{table_coeff_eta}).
\renewcommand{\arraystretch}{1.3}
\begin{table}[tb]
\begin{center}
\begin{tabular}{ c|c|c c c}
\multicolumn{2}{c|}{ \multirow{2}{*}{$\eta_\gamma^{(k)}$} } & \multicolumn{3}{c}{ $k$ }
\\
\cline{3-5}
\multicolumn{1}{c}{ } & & $0$ & $1$ & $2$
\\
\hline
\multirow{12}{*}{\rotatebox{90}{ $\gamma$ }} & $1$ & $0$ & $\pi^2/6$ & 0
\\
& $5/3$ & $-0.758981245$ & $1.517431001$ & $-2.60168706$
\\
& $2$ & $-1$ & $\pi^2/6$ & $-\pi^2/3$
\\
& $\alpha+2$ & $-1.10223102$ & $1.72183325$ & $-3.59345480$
\\
& $3$ & $-3/2$ &  $1/2+\pi^2/6$ &
\\
& $4$ & $-11/6$ & $1+\pi^2/6$ & 
\\
& $5$ & $-25/12$ & $35/24+\pi^2/6$ & 
\\
& $6$ & $-137/60$ & $45/24+\pi^2/6$ & 
\\
& $7$ & $-49/20$ & $203/90+\pi^2/6$ & 
\\
& $8$ & $-363/140$ & $469/180+\pi^2/6$ & 
\\
& $9$ & $-761/280$ & $29,531/10,080+\pi^2/6$ & 
\\
& $10$ & $-7,129/2,520$ & $6,515/2,016+\pi^2/6$ & 
\end{tabular}
\caption{
Values of the coefficients $\eta_\gamma^{(k)}$ 
calculated via the  equations  of appendix \ref{sec_appendix_sym_formula}.
\\
The calculations for $\gamma\in\mathbb{N}$ are analytical;   numerical otherwise.
For the specific $\eta_\alpha^{(k)}$ which depends on the value of $\alpha$,
that is of the effective interaction, we show here the result considering the SLy4 interaction ($\alpha=1/6$).
The $\eta_{i\in\mathbb{N}}^{(k)}$ are given 
 up to the $7^\mathrm{th}$ order
in the spin-orbit Taylor expansion (see text).
}
\label{table_coeff_eta}
\end{center}
\end{table}

Equation (\ref{eq_appendix_Imunu}) is an approximation for which the tiny error is $\sim\exp(-R/a)$.
One can observe that
\bea
\eta^{(0)}_{\gamma+1} - \eta^{(0)}_{\gamma}
=
- \frac{1}{\gamma}
\;\; ; \;\;
\eta^{(k)}_{\gamma+1} - \eta^{(k)}_{\gamma}
=
- \frac{k}{\gamma} \eta_{\gamma}^{(k-1)} \;\; {\scriptstyle (k>0)}
.
\label{eq_appendix_infinite_eta_k}
\eea 
\subsection{Expressions of a 3D integral as 1D integrals}
\label{sec_appendix_3Dto1D}
In this section we express the difference $\Delta I_{\gamma',\gamma} =I_{2,\gamma'}-I_{2,\gamma}$
as a sum of $1$-dimensional integrals.

The moments of the difference between two one-dimensional Fermi functions 
$F(x)=\left( 1+\e^{ x/a } \right)^{-1}$ 
to different powers $\gamma'$,$\gamma$ can be integrated as~\cite{krivine_math}
\bea
\int_{-\infty}^{+\infty}  x^k \Delta F_{\gamma',\gamma}(x) \mathrm dx
=
a^{k+1} \left( \eta^{(k)}_{\gamma'} - \eta^{(k)}_{\gamma}  \right)
,
\label{eq_appendix_infinite_int_F}
\eea
with $\Delta F_{\gamma',\gamma} = F^{\gamma'}-F^\gamma$.

Making the change of variable $x=r-R$, we can express the $3$-dimensional integral 
$I_{2,\gamma} =
\int \mathrm d\mathbf{r} F^{\gamma}(r)$
as a sum of three $1$-dimensional integrals of moments of Fermi functions $F(x)$:
\bea
I_{2,\gamma} &=& 4\pi \int_{-R}^{+\infty} (x+R)^2 F^\gamma(x) \mathrm dx +
4\pi \int_{-\infty}^{+\infty} (x+R)^2 F^\gamma(x) \mathrm dx \nonumber \\
&-& 4\pi \int_{-\infty}^{-R} (x+R)^2 F^\gamma(x) \mathrm dx
,
\label{eq_appendix_3D_changeVar}
\eea
where we have used the Chasles formula to  get integrals over the entire slab-space.
Assuming that the bulk is reached in the "negative" region, that is $F^\gamma(x<-R)=1$,
we can express the difference of two Fermi functions to different powers
\bea
\Delta I_{\gamma',\gamma} &=&
   4\pi R^2 \int_{-\infty}^{+\infty}    \Delta F_{\gamma',\gamma}(x) \mathrm dx
+  8\pi R   \int_{-\infty}^{+\infty} x  \Delta F_{\gamma',\gamma}(x) \mathrm dx \nonumber \\
&+&  4\pi    \int_{-\infty}^{+\infty} x^2 \Delta F_{\gamma',\gamma}(x) \mathrm dx
.
\label{eq_appendix_3Dto1D}
\eea
Because of the previous approximation,
we have spuriously inserted a bulk part in Eq.~(\ref{eq_appendix_3Dto1D}), 
but with the very tiny error $\sim  \Big( \exp(-\gamma' R/a) - \exp(-\gamma R/a) \Big)$.
Computing Eq.~(\ref{eq_appendix_3Dto1D}) with Eq.~(\ref{eq_appendix_infinite_int_F}) 
and expanding the radius parameter $R$ as a series of $(a/R_{HS})$ until the third order according to Eq.~(\ref{eq_sym_3D_R_1}),
we finally get at the third order in $(a/R_{HS})$:
\bea
\frac{\rho_{sat}}{3}\Delta I_{\gamma',\gamma} & =&
	 \left( \eta^{(0)}_{\gamma'} - \eta^{(0)}_{\gamma} \right)  \frac{a}{r_{sat}}  A^{2/3} \nonumber \\
	&+& 2    \left( \eta^{(1)}_{\gamma'} - \eta^{(1)}_{\gamma} \right)  \left( \frac{a}{r_{sat}} \right)^2  A^{1/3}\nonumber \\
	&+& \left(
			\left( \eta^{(2)}_{\gamma'} - \eta^{(2)}_{\gamma} \right)
			- \frac{2}{3}\pi^2   \left( \eta^{(0)}_{\gamma'} - \eta^{(0)}_{\gamma} \right)
			\right)
	\left( \frac{a}{r_{sat}} \right)^3 \nonumber \\
&+& O\left( \left(\frac{a}{r_{sat}}\right)^4 A^{-1/3} \right)
,
\label{eq_appendix_3D_deltaI_final}
\eea
with $r_{sat}=R_{HS}A^{-1/3} = \left( \frac{4}{3}\pi\rho_{sat}\right)^{-1/3}$.
Let us notice that we can also obtain equation~(\ref{eq_appendix_3D_deltaI_final}) using the general formula~(\ref{eq_appendix_Imunu}).
\section{Analytical expression for the surface energy}
\label{sec_appendix_sym_non_bulk}
We show in this section how equation~(\ref{eq_appendix_3D_deltaI_final}) allows 
to obtain an analytical formula for the symmetric local $E_{s}^L$ and non-local $E_{s}^{NL}$ surface energy
which lead to equations~(\ref{eq_sym_3D_Es_cl_L}), (\ref{eq_sym_3D_Es_cl_NL}) (\ref{eq_sym_coeff_L}) and (\ref{eq_sym_coeff_NL}).
We also detail the gaussian approximations as a function of $1$-dimensional integrals.
\subsection{The isoscalar local energy}
\label{sec_appendix_sym_L}
The surface local energy $E_{s}^L$ only depends on the density profile  $\rho(r)=\rho_{sat}F(r)$
through $h(\rho)=\sum_\gamma c_\gamma \rho^\gamma$ 
(see Eq.~(\ref{eq_energy_zeroth_loc_sym}) for the values of $\gamma$, $c_\gamma$), such that
\bea
E_{s}^{L}
=
\int \mathrm d\mathbf{r} \left\{  h(\rho) - \frac{h(\rho_{sat})}{\rho_{sat}} \rho \right\}
=
 \sum_\gamma c_\gamma\rho_{sat}^\gamma 
	\Delta I_{\gamma,1}
.
\label{eq_appendix_sym_3D_L}
\eea
Computing with the equation~(\ref{eq_appendix_3D_deltaI_final}) 
for the $\gamma$-values of $h(\rho)$
($\gamma=5/3$, $8/3$, $2$ and $(\alpha+2)$), we obtain
\begin{widetext}
\begin{align}
E_{s}^{L} & =
3
\left\{
	\frac{\hbar^2}{2m}\left( \frac{3\pi^2}{2} \right)^{2/3} \frac{3}{5} \rho_{sat}^{2/3} 
				\left[
					 \eta^{(0)}_{5/3} 
					 +  \kappa\rho_{sat} 
					 		\eta^{(0)}_{8/3}
				\right]
	- C_0 \rho_{sat}
	+ C_3 \rho^{\alpha+1}_{sat}  \eta^{(0)}_{\alpha+2}
\right\}
	\frac{a}{r_{sat}} A^{2/3}
\nonumber \\	 & \hphantom{  \simeq }
+
6
\left\{
	\frac{\hbar^2}{2m}\left( \frac{3\pi^2}{2} \right)^{2/3} \frac{3}{5} \rho_{sat}^{2/3} 
				\left[
				 	  \eta^{(1)}_{5/3}
				 	+ \kappa\rho_{sat} \eta^{(1)}_{8/3}	
					- \frac{\pi^2}{6} \frac{m}{m^*_{sat}} 
				\right]	
+ C_3 \rho_{sat}^{\alpha+1} 
	\left(
		\eta^{(1)}_{\alpha+2}
		- \frac{\pi^2}{6}
	\right)
\right\}
	\left( \frac{a}{r_{sat}}	\right)^2  A^{1/3} 
\nonumber \\	 & \hphantom{  \simeq }
+
3
\left\{
	\frac{\hbar^2}{2m} \left( \frac{3\pi^2}{2} \right)^{2/3} \frac{3}{5} \rho_{sat}^{2/3} 
		    	\left[	
					\eta^{(2)}_{5/3}
					- \frac{2 \pi^2}{3} \eta^{(0)}_{5/3}
					+ \kappa\rho_{sat}
						\left(	
							\eta^{(2)}_{8/3}
							- \frac{2 \pi^2}{3} \eta^{(0)}_{8/3}
						\right)
			\right]
\right.
\nonumber \\ & \hphantom{ \simeq + \Bigg\{ }
\left.  \vphantom{ \left( \frac{3\pi^2}{2} \right)^{2/3} }
	+ \frac{\pi^2}{3} C_0 \rho_{sat}
	+ C_3 \rho_{sat}^{\alpha+1} 
		\left(	
			\eta^{(2)}_{\alpha+2}
			- \frac{2 \pi^2}{3} \eta^{(0)}_{\alpha+2}
		\right)
\right\}
	\left( \frac{a}{r_{sat}}	\right)^3
+ O\left( \left(\frac{a}{r_{sat}}\right)^4 A^{-1/3} \right)
,	
\label{eq_appendix_sym_3D_nb_L}
\end{align}
\end{widetext}
where the values of  
$\eta^{(k)}_{\gamma}$ are given in table \ref{table_coeff_eta}. 
Using (\ref{eq_appendix_infinite_eta_k})
which gives a relation between $\eta^{(k)}_{8/3}$ and $\eta^{(k)}_{5/3}$,
we get the local energy $E_{nb}^L$ as a function of $\eta^{(k)}_{5/3}$ and $\eta^{(k)}_{\alpha+2}$ only ($k=0,1,2$)
(Eqs.~(\ref{eq_sym_3D_Es_cl_L}), (\ref{eq_sym_3D_Es_cl_NL}) and (\ref{eq_sym_coeff_L})).
\subsection{The symmetric non-local energy}
\label{sec_appendix_sym_NL}
The non-local energy $E_{s}^{NL}$ is the integration of a quadratic function in the density gradient
such that we can put it on the form
$\sum_\gamma c_\gamma \left(\nabla\rho\right)^2 \rho^{\gamma-2}$
(see Eq.~(\ref{eq_sym_3D_Enb_NL}) for the values of $\gamma$, $c_\gamma$).
Expressing the Fermi gradient function as follows
\bea
\nabla\rho(r) = 
\rho_{sat} \nabla F(r)
\;\; ; \;\;
\nabla F(r) =
\frac{1}{a} \left( F^2(r) - F(r) \right)
,
\label{eq_sym_3D_rho_prime}
\eea
we can write
\begin{align}
E_{s}^{NL} & =
\int \mathrm d\mathbf{r} \sum_\gamma c_\gamma \left( \nabla\rho  \right)^2 \rho^{\gamma-2}\nonumber \\
&= 
\frac{1}{a^2}
\sum_\gamma c_\gamma \rho_{sat}^{\gamma}   \int \mathrm d\mathbf{r} 
	\Big[
	\Big( F^{\gamma+2} - F^{\gamma+1} \Big) - \Big( F^{\gamma+1} - F^{\gamma} \Big)
	\Big]
\nonumber \\
& =
\sum_\gamma c_\gamma \rho_{sat}^{\gamma} \Big[ \Delta I_{\gamma+2,\gamma+1} - \Delta I_{\gamma+1,\gamma} \Big]
.
\label{eq_appendix_sym_3D_nb_NL}
\end{align}
For $\gamma \geq 1$, 
using the recursion relation (\ref{eq_appendix_infinite_eta_k}), we have
\begin{align}
\eta^{(0)}_{\gamma+2} - \eta^{(0)}_{\gamma+1} 
-	\left( 
		\eta^{(0)}_{\gamma+1}  - \eta^{(0)}_{\gamma} 
	\right)
&  = 
 \frac{1}{\gamma \left( \gamma+1 \right)} ,
\nonumber \\
\eta^{(1)}_{\gamma+2} - \eta^{(1)}_{\gamma+1} 
-	\left( 
		\eta^{(1)}_{\gamma+1}  - \eta^{(1)}_{\gamma} 
	\right)
&  = 
 \frac{1}{\gamma \left( \gamma+1 \right)} \left( \eta^{(0)}_{\gamma} + 1 \right) ,
\nonumber \\
\eta^{(2)}_{\gamma+2} - \eta^{(2)}_{\gamma+1} 
-	\left( 
		\eta^{(2)}_{\gamma+1}  - \eta^{(2)}_{\gamma} 
	\right)
&  = 
 \frac{2}{\gamma \left( \gamma+1 \right)} \left(  \eta^{(1)}_{\gamma} + \eta^{(0)}_{\gamma}  \right)
,
\label{eq_appendix_eta_k_2_1}
\end{align}
which allows simplifying the expression of $E_{s}^{NL}$
once we have computed Eq.~(\ref{eq_appendix_sym_3D_nb_NL})
with Eq.~(\ref{eq_appendix_3D_deltaI_final}):
\begin{align}
E_{s}^{NL}
& =
\frac{1}{a^2}
\sum_\gamma c_\gamma
\rho_{sat}^{\gamma-1}
\frac{1}{\gamma \left( \gamma +1 \right)} \nonumber \\
&\left\{
		3\frac{a}{r_{sat}} A^{2/3}
		+ 6 \left[
					\eta^{(0)}_{\gamma} + 1
			\right]
			\left( \frac{a}{r_{sat}} \right)^2 A^{1/3}
\right. \nonumber \\ & \left.
	\hphantom{ \frac{1}{a^2}   \Big\{ }
  + 6 \left[
					\eta^{(1)}_{\gamma}  
					+ \eta^{(0)}_{\gamma}
					- \frac{\pi^2}{3}
			\right]
			\left( \frac{a}{r_{sat}} \right)^3
\right\}
\nonumber \\
& + O\left( \left(\frac{a}{r_{sat}}\right)^4 A^{-1/3} \right)
.
\label{eq_appendix_sym_3D_nb_NL_simplified}
\end{align}
Looking at the definition of the non-local energy $E_{s}^{NL}$ Eq.~(\ref{eq_sym_3D_Enb_NL}),
one can see that there are terms $\propto f^{-1}=(1+\kappa\rho)^{-1}$. 
In order to have an expression in the form of Eq.~(\ref{eq_appendix_sym_3D_nb_NL_simplified}),
we need to make a Taylor expansion, such that $f^{-1} = \sum_{i=0} (-1)^i (\kappa\rho)^i$. 
Then we can straightforwardly compute the non-local energy with Eq.~(\ref{eq_appendix_sym_3D_nb_NL_simplified})
(with $\gamma=1$, $2$, $i+2$ and $i+3$)
to obtain Eqs.~(\ref{eq_sym_3D_Es_cl_L}), (\ref{eq_sym_3D_Es_cl_NL}) and (\ref{eq_sym_coeff_NL}).
\subsection{The isovector energy}
\label{sec_appendix_asym}
In this section we develop the $3$-dimensional gaussian $\mathcal{G}(r)$ integral
\bea
E_G &=& 4 \pi \int_0^\infty \mathrm dr r^2 \mathcal{G}(r) ,
\nonumber \\ 
&=& 4 \pi \int_0^\infty \mathrm dr r^2 \mathcal{A} \exp\left( - \frac{(r-R_M)^2}{2\sigma^2} \right)
\label{eq_appendix_Gauss_def}
\eea
as $1$-dimensional integrals in order to obtain the isovector surface energy as a function of the nucleus mass
and of the effective interaction parameters.
As for the symmetric energy, we make the variable change $x=r-R_M$:
\bea
E_G &=& 
4 \pi \mathcal{A} \int_{-\infty}^{+\infty} \mathrm dx \left(x+R_M\right)^2 \exp\left( - \frac{x^2}{2\sigma^2} \right)
\nonumber \\ &-&
4 \pi \mathcal{A} \int_{-\infty}^{-R_M}    \mathrm dx \left(x+R_M\right)^2  \exp\left( - \frac{x^2}{2\sigma^2} \right)
.
\label{eq_appendix_Gauss_1D}
\eea
Since we are interested in the surface energy, 
we  assume that the gaussian $\mathcal{G}(r)$ is zero at the center of the nucleus,
such that the second integral in Eq.~(\ref{eq_appendix_Gauss_1D}) is negligible 
with an accuracy $\sim \exp\left( -R_M^2/(2\sigma^2) \right)$.
Then integrating the gaussian moments straightforwardly lead to
\bea
E_G  =   2\left( 2\pi \right)^{3/2} \sigma \mathcal{A}  \left( R_M^2 + \sigma^2 \right) 
,
\label{eq_appendix_Gauss_integr_gen}
\eea
where we have used the expression of the variance as a function of the energy density second derivatives
(see section~\ref{sec_asym_Gauss_apprx}).
To have $E_G$ as a function of the mass, we just need to express the gaussian maximum position $R_M$
as a function of $A$.
If we assume $R_M=R$, it reads, using Eq.~(\ref{eq_sym_3D_R_1}):
\bea
E_G 
& =&  2\left( 2\pi \right)^{3/2} \sigma \mathcal{A}
 r_{sat}^2 \nonumber \\
&&		\left[  
		A^{2/3} 
		+ \frac{\sigma^2}{r_{sat}^2}
		- \frac{2\pi^2}{3} \left( \frac{a}{r_{sat}} \right)^2
		+ O\left( \left(\frac{a}{r_{0}}\right)^4 A^{-2/3} \right)
		\right] \nonumber \\
\label{eq_appendix_Gauss_integr_R}
\eea
In the general case, if we define $\Delta R = R_M - R$, we find additional terms, especially curvature:
\begin{align}
E_G 
& = 2\left( 2\pi \right)^{3/2} \sigma \mathcal{A}
	r_{sat}^2 \nonumber \\
&		\left[ 
		A^{2/3} 
		+ 2 \frac{\Delta R}{r_{sat}} A^{1/3}
		+ \frac{\sigma^2}{r_{sat}^2}
		- \frac{2\pi^2}{3} \left( \frac{a}{r_{sat}} \right)^2
		+ \left(  \frac{\Delta R}{r_{sat}} \right)^2 
\right.
\nonumber \\ & 
\hphantom{ =  2   \Big[  } 
\left.
		- \frac{2\pi^2}{3} \frac{\Delta R}{r_{sat}} 	\left(\frac{a}{r_{0}}\right)^2 A^{-1/3}
		+  O\left( \left(\frac{a}{r_{0}}\right)^4 A^{-2/3} \right)
\right]
.
\label{eq_appendix_Gauss_integr_DeltaR}
\end{align}


\begin{thebibliography}{99}
\bibitem{footnote1} Notice that this decomposition is not the same as in Eq.(\ref{tau_loc}),(\ref{tau_nloc}).
\bibitem{footnote2} Notice that extreme values of the order $\delta\approx 0.4$ are obtained only for the lightest nuclei, not 
considered in this paper.
\end{thebibliography}
\end{document}